\pdfoutput=1
\documentclass[longauth]{aa}
\usepackage{graphicx}
\usepackage{xcolor}
\usepackage{array}
\newcolumntype{P}[1]{>{\centering\arraybackslash}p{#1}}
\usepackage[bookmarks=false,colorlinks=true,linkcolor=black,citecolor=cyan,filecolor=black,urlcolor=cyan]{hyperref}
\usepackage{amsmath}
\usepackage{multirow,tabularx}

\usepackage{makecell}
\usepackage{newtxtext,newtxmath}

\usepackage[T1]{fontenc}
\usepackage{ae,aecompl}

\usepackage{xspace} 
\usepackage{siunitx} 
\usepackage{bm}
\usepackage{syntonly}
\usepackage{booktabs}
\usepackage{threeparttable}
\usepackage{pdflscape}
\usepackage{longtable}
\usepackage{makecell}

\usepackage{ltablex}
\usepackage{boxhandler}
\usepackage{graphbox}

\graphicspath{{./figures/}}

\begin{document} 

   \title{The eROSITA Final Equatorial Depth Survey (eFEDS):}
   
   \subtitle{SDSS spectroscopic observations of X-ray sources}

    \titlerunning{eFEDS SDSS spectroscopic follow-up} 
    

   \author{C. Aydar\inst{1, 2}\thanks{\href{mailto:caydar@mpe.mpg.de}{caydar@mpe.mpg.de}} \and
           A. Merloni\inst{1} \and
            T. Dwelly\inst{1} \and
            J. Comparat\inst{1} \and
            M. Salvato\inst{1} \and
            J. Buchner\inst{1} \and
            M. Brusa\inst{3, 4} \and
            T. Liu\inst{5} \and
            J. Wolf\inst{6} \and
            S.~F. Anderson\inst{7} \and
            C.~P. Andonie\inst{1} \and
            F.~E. Bauer\inst{8} \and
            M. R. Blanton\inst{9} \and
            W.~N. Brandt\inst{10, 11, 12} \and
            Y. D\'iaz\inst{13} \and
            L. Hern\'andez-Garc\'ia\inst{14, 15, 16} \and
            D.-W. Kim\inst{17} \and
            T. Miyaji\inst{18} \and
            S. Morrison\inst{19} \and
            B. Musiimenta\inst{3, 4} \and
            C.~A. Negrete\inst{20} \and
            Q. Ni\inst{1} \and
            C. Ricci\inst{13, 21} \and
            D.~P. Schneider\inst{10, 11} \and
            A. Schwope\inst{22} \and
            Y. Shen\inst{19, 23} \and
            S.~G.~H. Waddell\inst{1, 24} \and
            R. Arcodia\inst{1, 25} \and
            D. Bizyaev\inst{26, 27} \and
            J.~N. Burchett\inst{28} \and
            P. Chakraborty\inst{17} \and
            K. Covey\inst{29} \and
            B.~T. G\"ansicke\inst{30} \and
            A. Georgakakis\inst{31} \and
            P.~J. Green\inst{17} \and
            H. Ibarra\inst{20} \and
            J. Ider-Chitham\inst{1} \and
            A.~M. Koekemoer\inst{32} \and
            J.~A. Kollmeier\inst{33, 34} \and
            M. Krumpe\inst{22} \and
            G. Lamer\inst{22} \and
            A. Malyali\inst{1} \and
            K. Nandra\inst{1} \and
            K. Pan\inst{26} \and
            C.~R. Pizarro\inst{22} \and
            J. S\'anchez-Gallego\inst{35} \and
            J.~R. Trump\inst{36} \and
            T. Urrutia\inst{22}
            }

   \institute{Max Planck Institute for Extraterrestrial Physics, Gie{\ss}enbachstra{\ss}e 1, 85748 Garching, Germany
        \and
            Excellence Cluster ORIGINS, Boltzmannstrasse 2, D-85748 Garching, Germany    
        \and
            Dipartimento di Fisica e Astronomia, Universit\`a di Bologna, Via Gobetti 93/2, I-40129 Bologna, Italy 
        \and
            INAF – Osservatorio di Astrofisica e Scienza dello Spazio di Bologna, Via Gobetti 93/3, I-40129 Bologna, Italy 
        \and
            Department of Astronomy, University of Science and Technology of China, Hefei 230026, People's Republic of China 
        \and
            Max Planck Institut f\"ur Astronomie, K\"onigstuhl 17, 69117 Heidelberg, Germany 
        \and
            Department of Astronomy, University of Washington, Box 351580, Seattle, WA 98195, USA 
        \and
            Instituto de Alta Investigaci{\'{o}}n, Universidad de Tarapac{\'{a}}, Casilla 7D, Arica, Chile 
        \and
            Center for Cosmology and Particle Physics, Department of Physics, 726 Broadway, Room 1005, New York University, New York, NY 10003, USA 
        \and
            Department of Astronomy \& Astrophysics, 525 Davey Lab, The Pennsylvania State University, University Park, PA 16802, USAA 
        \and
            Institute for Gravitation and the Cosmos, The Pennsylvania State University, University Park, PA 16802, USA 
        \and
            Department of Physics, 104 Davey Laboratory, The Pennsylvania State University, University Park, PA 16802, USA 
        \and
            Instituto de Estudios Astrof\'isicos, Facultad de Ingenier\'ia y Ciencias, Universidad Diego Portales, Av. Ej\'ercito Libertador 441, Santiago, Chile 
        \and
            Millennium Nucleus on Transversal Research and Technology to Explore Supermassive Black Holes (TITANS), Gran Breta\~na 1111, Playa Ancha, Valpara\'iso, Chile 
        \and
            Millennium Institute of Astrophysics (MAS), Nuncio Monse\~nor S\'otero Sanz 100, Providencia, Santiago, Chile 
        \and
            Instituto de F\'isica y Astronom\'ia, Facultad de Ciencias, Universidad de Valpara\'iso, Gran Breta\~na 1111, Playa Ancha, Valpara\'iso, Chile 
        \and
            Center for Astrophysics, Harvard \& Smithsonian, Cambridge, MA 02138, USA 
        \and 
            Instituto de Astronom\'ia, Universidad Nacional Aut\'onoma de M\'exico Campus Ensenada, Km 107,  Carret. Tijuana-Ensenada, 22860, Ensenada, Mexico 
        \and
            Department of Astronomy, University of Illinois Urbana-Champaign, Urbana, IL 61801, USA 
        \and    
            Instituto de Astronom\'ia, Universidad Nacional Aut\'onoma de M\'exico, A.P. 70-264, 04510 M\'exico D. F., Mexico 
        \and
            Kavli Institute for Astronomy and Astrophysics, Peking University, Beijing 100871, China 
        \and
            Leibniz-Institut f\"ur Astrophysik Potsdam (AIP), An der Sternwarte 16, 14482 Potsdam, Germany 
        \and
            National Center for Supercomputing Applications, University of Illinois Urbana-Champaign, Urbana, IL 61801, USA 
        \and
            Trottier Space Institute \& Department of Physics, McGill University, 3600 rue University, Montreal, QC, H3A 2T8, Canada 
        \and
            Kavli Institute for Astrophysics and Space Research, Massachusetts Institute of Technology, Cambridge, MA 02139, USA 
        \and
            Apache Point Observatory and New Mexico State University, P.O. Box 59, Sunspot, NM, 88349-0059, USA 
        \and
            Sternberg Astronomical Institute, Moscow State University, Moscow, Russia 
        \and 
            Department of Astronomy, New Mexico State University, Las Cruces, NM 88003, USA 
        \and
            Department of Physics \& Astronomy, Western Washington University, 516 High Street, Bellingham, WA 98225, USA 
        \and 
            Department of Physics, University of Warwick, Coventry CV4 7AL, UK 
        \and
            Institute for Astronomy \& Astrophysics, National Observatory of Athens, V. Paulou \& I. Metaxa, 11532, Greece 
        \and
            Space Telescope Science Institute, 3700 San Martin Drive, Baltimore, MD 21218, USA 
        \and
            The Observatories of the Carnegie Institution for Science, 813 Santa Barbara Street, Pasadena, CA 91101, USA 
        \and
            Canadian Institute for Theoretical Astrophysics, University of Toronto, Toronto, ON M5S 98H, Canada 
        \and
            Department of Astronomy, University of Washington, Box 351580, Seattle, WA 98195, USA 
        \and
            Department of Physics, 196 Auditorium Road, Unit 3046, University of Connecticut, Storrs, CT 06269 USA 
        }
           
   \date{}

  \abstract
  {We present one of the largest uniform optical spectroscopic surveys of X-ray selected sources to date that were observed as a pilot study for the Black Hole Mapper (BHM) survey.
  The BHM program of the Sloan Digital Sky Survey (SDSS)-V is designed to provide optical spectra for hundreds of thousands of X-ray selected sources from the SRG/eROSITA all-sky survey.
  This significantly improves our ability to classify and characterise the physical properties of large statistical populations of X-ray emitting objects. 
  Our sample consists of 13\,079 sources in the eROSITA eFEDS performance verification field, 12\,011 of which provide reliable redshifts from $0\lesssim z\leq 5.8$.
  The vast majority of these objects were detected as point-like sources (X-ray flux limit $F_{\rm 0.5-2 keV}\gtrsim 6.5\times 10^{-15}$ erg/s/cm$^2$) and were observed for about 20 years with fibre-fed SDSS spectrographs.
  After including all available redshift information for the eFEDS sources from the dedicated SDSS-V plate programme and archival data, we visually inspected the SDSS optical spectra to verify the reliability of these redshift measurements and the performance of the SDSS pipeline.
  The visual inspection allowed us to recover reliable redshifts (for 99\% of the spectra with a signal-to-noise ratio of $>2$ ) and to assign classes to the sources, and we confirm that the vast majority of our sample consists of active galactic nuclei (AGNs).
  Only $\sim3\%$ of the eFEDS/SDSS sources are Galactic objects.
  We analysed the completeness and purity of the spectroscopic redshift catalogue, in which the spectroscopic completeness increases from $48\%$ (full sample) to $81\%$ for a cleaner, brighter ($r_{\rm AB}<21.38$) sample that we defined by considering a high X-ray detection likelihood, a reliable counterpart association, and an optimal sky coverage.  
  We also show the diversity of the optical spectra of the X-ray selected AGNs and provide spectral stacks with a high signal-to-noise ratio in various sub-samples with different redshift and optical broad-band colours.
  Our AGN sample contains optical spectra of (broad-line) quasars, narrow-line galaxies, and optically passive galaxies. 
  It is considerably diverse in its colours and in its levels of nuclear obscuration.  
  }
 \keywords{galaxies: active - galaxies: evolution - quasars: emission lines - X-rays: galaxies - techniques: optical spectroscopy}

\maketitle

\section{Introduction}  

To properly understand the evolution of any class of astrophysical objects through cosmic time, it is necessary not only to study individual sources in detail, but also to consider their behaviour as a (statistical) population.
Demographic studies further our understanding of the distinct phases in the evolution of a type of object because extensive samples can provide enough statistics to show hidden trends.
However, to build reliable statistical samples, large collaborations must be involved in a systematic approach to observe numerous sources over the entire sky.
This paper introduces a sample of point-like X-ray detected sources that spans a wide redshift range ($0\le z <5.8$), for which optical spectra were obtained systematically with the Sloan Digital Survey (SDSS) telescope, supplemented by spectroscopic redshift information from the literature.

The first step in building this catalogue involved the detection of a large sample of X-ray sources \citep{Brandt2005}.
X-ray astronomy has developed more recently than optical astronomy \citep{Elvis2020}, and the first major wide-area survey conducted in X-rays was the ROSAT \citep{Truemper1982} All-Sky Survey in 1990.
ROSAT observed the sky in the $0.1-2.4$ keV band and detected more than $10^5$ unique sources. This sample is more than 100 times larger than previous X-ray surveys \citep{Boller2016}. 
Since then, \textit{XMM-Newton} \citep{Jansen2001_XMM} and \textit{Chandra} \citep{Weisskopf2002_Chandra} have been observing in the X-rays with their larger collecting area and better spatial resolution.
Although these observatories increased the number of known X-ray emitting sources significantly, their small fields of view were not designed for an all-sky survey. They therefore covered limited patches of the sky (the overall coverage of \textit{XMM-Newton} with 4XMM is 1\,383 deg$^{2}$ and that of \textit{Chandra} with CSC2 is 783 deg$^{2}$; see \citealt{Merloni2024} and references therein).
The different settings for each observation also make any statistical population analysis more challenging because it is a complex task to account for the different calibrations, exposure times, depths, and sensitivities.

The extended ROentgen Survey with an Imaging Telescope Array \citep[eROSITA,][]{Predehl2021} on board the Spectrum Roentgen Gamma satellite \citep[SRG,][]{Sunyaev2021} is the most powerful wide-field X-ray survey telescope to date and was designed to provide X-ray spectroscopy and imaging of the entire sky.
It was optimised to deliver a large effective area and field of view in the soft X-ray band while scanning the whole sky and to have an angular resolution that is good enough to distinguish point-like and large extended X-ray sources such as clusters of galaxies.
The first release of the eROSITA all-sky survey data \citep{Merloni2024} increased the number of known X-ray sources in the literature by $>60\%$. 
It detected almost one million objects, including stars, compact objects, galaxies, active galactic nuclei (AGNs), and clusters of galaxies.

Access to information at longer wavelengths allows the interpretation of the data for any X-ray catalogue.
In particular, the information provided by optical spectroscopy (e.g. the redshift) permits a fit of the observed X-ray spectra with a correct model, and it also allows us to fix the basic source distance and luminosity for each object.
The SDSS \citep{York2000} has been providing astronomical optical-to-IR data for more than two decades.
This laid the basis for several ground-breaking demographic studies of millions of sources from different astronomical classes in the Milky Way and beyond (see \citealt{Ahumada2020_sdss_dr16} for the 16th data release and \citealt{Almeida2023_sdss_dr18} for the 18th data release).

The multi-object spectroscopic capability of the SDSS (a few tens up to a few hundred objects per square degree can be observed simultaneously) matches the expected sky density of the X-ray sources that are detected at the typical depth of the eROSITA all-sky survey well \citep{Merloni2012}.
The programme SPectroscopic IDentification of eROSITA Sources \citep[SPIDERS,][]{Dwelly2017, Comparat2020} was devised more than a decade ago, with the ultimate goal of providing SDSS optical spectra for a large number of X-ray sources that were detected by eROSITA so that their highly energetic processes (e.g. matter accretion, hot gas emission) could be studied, together with some of the physical properties that are often obtained from the emission and absorption lines in the optical-to-IR domain, such as density, metallicity, the ionisation parameter, and the contribution from star formation \citep[e.g.][]{Kewley2019}.

In this paper, we present the value-added catalogue that was produced based on the optical spectra obtained from X-ray sources in the eFEDS field \citep[eROSITA Final Equatorial Depth Survey,][]{Brunner2022}.
The eFEDS observations covered $\sim$140\,deg$^2$ and were carried out as part of eROSITA performance verification to demonstrate that eROSITA was able to meet its all-sky survey science goals. As a pilot survey, eFEDS X-ray observations in this field are $\sim40\%$ deeper than the planned all-sky survey \citep{Predehl2021, Brunner2022, Merloni2024}.
This allows more detailed studies of each object and the detection of a larger sample of faint objects.
The eFEDS field location was selected to exploit existing observations at other wavelengths, which enabled the timely identification of the X-ray counterparts \citep{Salvato2022}.

At the time of its completion (November 2019), eFEDS was the X-ray survey medium-depth field with the largest number of detected sources in a contiguous footprint \citep[see Fig.~14 from][]{Brunner2022}. 
Since the eFEDS data were published, many studies on the X-ray \citep[e.g.][]{Brunner2022, Liu2022, Comparat2023, Schwope2024, Waddell2024, Nandra2025} and multi-wavelength \citep[e.g.][]{Salvato2022, Bulbul2022, Toba2022, Schneider2022, Klein2022, Li2024_HSC, Igo2024} properties of the sample were published.

One great advantage of this large dataset is that it was obtained with only one instrument for each wavelength range.
This instrumental uniformity allows a more consistent statistical comparison of the objects because technical differences in calibration and the observing method do not have to be considered.
A large and uniform sample allows the analysis of the parameter space through different bins of the main properties that drive the evolution of any class of astrophysical object.
One possible technique that empowers this type of analysis is stacking. We used it to analyse the optical spectra of AGNs and quasars (QSOs)\footnote{In this paper, we refer to AGNs and QSOs interchangeably. Historically, QSOs can refer to ``bright AGNs'', although there is no standard luminosity definition \citep[see e.g.][]{Peterson1997}.} because they are the most common population of X-ray point-like sources that were found at the depth of eFEDS \citep[][]{Menzel2016, Liu2022}.
Some of the key measurable features that influence the AGN evolution and can be studied with this catalogue are the luminosity, the black hole mass, and the fraction of the contribution from the host galaxy to the overall emission \citep[e.g.][]{Brandt2015, Merloni2016}, and the level of intervening obscuration by gas and dust.
These measured quantities allowed us to address questions regarding the growth of supermassive black holes \citep[SMBH, e.g.][]{Brandt2005, Alexander2012}, the role of obscuration in observations of AGNs \citep[e.g.][]{Hickox_Alexander_2018}, the coevolution of SMBHs and their host galaxies \citep[e.g.][]{Hopkins2008, Hickox2009, Kormendy2013}, and the comprehension of AGNs as the standard unification model and an evolutionary model \citep[e.g.][]{Antonucci1993, Netzer2015}.
For non-AGNs, the eFEDS catalogue also provides large samples that allow us to perform reliable statistical studies.

This manuscript is structured as follows.
Section~\ref{sec:data} presents the datasets of the eFEDS field, considering the X-ray data of eROSITA, the optical-to-IR spectroscopic data of SDSS, and the collection of spectroscopic redshifts obtained from the cross-match with other surveys that have been conducted in the eFEDS field.
Section~\ref{sec:methods} describes the visual inspection process with which we confirmed the SDSS pipeline attribution of the redshift to the point-like X-ray detected sources and the method we followed to compile the spectroscopic redshifts from the available literature sources.
Section~\ref{sec:results:general} presents the analysis of the different classes of objects we found in the SDSS/eFEDS catalogue and discusses some of the statistical properties of the spectroscopic sample. The focus lies on the completeness and purity of the catalogue.
A qualitative analysis of the diversity of the extragalactic point-like sources is presented in Sect.~\ref{sec:results:xray_pointlike}.
We focus on AGNs and present stacks with a high signal-to-noise ratio as a function of redshift and optical colour.
Section~\ref{sec:conclusions} summarises the catalogue and provides an outlook of the ongoing and future observations of eROSITA targets with SDSS-V.
We also make the catalogues that we compiled and used for the plots available. 
Their data models are described in the appendix (see Appendix \ref{sec:appendix:catalogues} and \ref{sec:data_model_efeds_speccomp}).
Throughout this manuscript, we adopt a flat $\Lambda$CDM cosmology with $\Omega_{\rm{M}} = 0.3$ and H$_0 = 70$ km s$^{-1}$ Mpc$^{-1}$.
Unless stated otherwise, the photometric magnitudes are given in the AB system \citep{OkeGunn1983}.
We use the term spec-z for a catalogue entry that provides a redshift measurement (and optionally, a broad classification) that was originally derived from (usually optical) spectroscopy.

\section{Data}
\label{sec:data}

In this Section, we present the observations of the eFEDS field.
We describe the eROSITA X-ray observations, the SDSS optical spectroscopic observations, and the compilation of redshifts for the detected sources considering multi-wavelength data from several surveys.

\subsection{The eFEDS field: X-ray observations and multi-wavelength counterparts}
\label{sec:eFEDS_field}

The eROSITA X-ray telescope operates in the $0.2-8.0$ keV band, with the main sample comprising the objects detected in the $0.2-2.3$ keV band \citep{Brunner2022}, where the instrument is most sensitive \citep[see Fig.~10 from][]{Predehl2021}.
To verify the calibration and performance of eROSITA before starting the all-sky scanning operations, eROSITA observed a small region of the sky with long exposures. 
The eROSITA Final Equatorial Depth Survey (eFEDS) field comprises a region of $\sim$140 deg$^2$ centred at RA$=136^{\rm{o}}$ and Dec$=+2^{\rm{o}}$ (see Fig.~\ref{fig:radec}).
The eFEDS field was observed by eROSITA for about 360 ks (100 hours) in total between November 3 and 7, 2019, performing uniform exposures of $\sim2.2$ ks ($\sim1.2$ ks after correcting for telescope vignetting), $\sim40\%$ deeper than the originally planned observations of the all-sky survey in 8 scans.
This procedure produced a source catalogue with limiting flux of $F_{0.5-2\ \rm{keV}} \sim 6.5 \times 10^{-15}\ \rm{erg}\ \rm{s}^{-1}\ \rm{cm}^{-2}$, sufficiently deep to allow for the detection of large statistical samples of different types of astronomical objects (e.g. stars, compact objects, galaxies, clusters of galaxies, and AGNs).
The limiting flux of eFEDS reaches $\sim15-20$ times fainter sources than the one achieved by the previous X-ray all-sky survey, ROSAT, for a similar soft X-ray band \citep[$0.1-2.4$ keV,][]{Boller2016}.

The eROSITA/eFEDS main sample includes 27\,368 point-like, and 542 extended X-ray sources \citep[total 27\,910 sources,][]{Brunner2022}, following a classification entirely based on the detected X-ray morphology.
Out of these extended sources, 102 were confirmed to be galaxy clusters in \citet{LiuAng2022}.
Among the point-like sources, a small number of misclassified galaxy clusters were identified in \citet[][see Sect.~\ref{sec:results:xray_pointlike}]{Bulbul2022}.

\subsection{SDSS spectroscopic observations of the eFEDS field}
\label{sec:SDSS_data}

The $2.5$ m telescope located at the Apache Point Observatory \citep{Gunn2006} observed a large portion of the Northern sky with the fibre system of the SDSS and the Baryon Oscillation Spectroscopic Survey \citep[BOSS,][]{Gunn2006, Dawson2013_boss, Smee2013}, considering different targeting programs to obtain significant samples of several types of objects.
This instrument is well suited to the task because of its high multiplex, wide field-of-view, broad bandpass, and high efficiency. 
We have concentrated spectroscopic follow-up efforts on the part of the X-ray source population having optical counterparts with magnitude $16<r<22$ AB. 

This section provides a brief recap of the spectroscopic observations obtained in the eFEDS field over several generations of the SDSS project.
We can break this down into the following four phases most relevant to eFEDS: 
\begin{itemize}
\item[i)] The SDSS-I, II galaxy and QSO legacy redshift survey \citep[][]{York2000,Strauss2002,Eisenstein2001,Strauss2002,Abazajian2009}, 
\item[ii)] The SDSS-III Baryon Oscillation Spectroscopic Survey \citep[BOSS,][]{Eisenstein2011,Dawson2013_boss},
\item[iii)] The SDSS-IV/eFEDS special plate programme \citep[DR17, ][their Sect.~7.5]{Blanton2017, Abdurrouf_2022_sdss_dr17}, 
\item[iv)] The SDSS-V/eFEDS special plate programme \citep[DR18,][their Sect.~9.1]{Almeida2023_sdss_dr18}
\end{itemize}

The SDSS-I, II, and III spectroscopy was obtained before the launch of SRG/eROSITA, but has serendipitously observed many objects that were discovered later to be optical counterparts to eFEDS X-ray sources.
For these, we have used data products distributed as part of SDSS~DR16 \citep[][]{Ahumada2020_sdss_dr16}.  

\begin{figure*}[t]
\centering
\includegraphics[width=\textwidth]{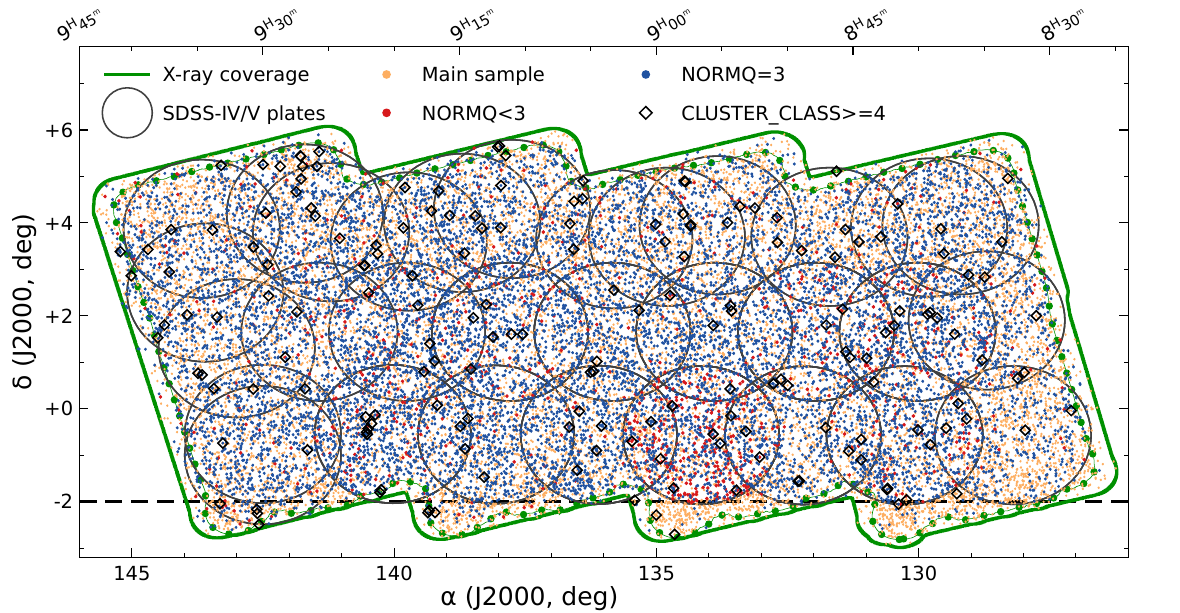}
  \caption{Layout of the eFEDS field for the SPIDERS targeting and observations.
    The eROSITA X-ray coverage of the eFEDS field is shown in green, with the region with $>$500\,s of effective exposure depth shown with connected green points.
    SDSS-V plates are indicated as large black circles.
    The eFEDS Main X-ray sample is marked with small orange dots.
    The main sample sources having reliable spectroscopic redshifts (\texttt{NORMQ}$=$3, see Table \ref{tab:vi_grading}) are shown in dark blue, with red points indicating cases with unreliable redshift estimates (\texttt{NORMQ}$<$3).
    The black diamonds indicate the objects that were detected in the X-ray as point-like sources, but which have counterparts suggesting that the source is likely associated with a galaxy cluster (\texttt{CLUSTER\_CLASS}$\geq$4 from \citealt{Salvato2022}).
    The horizontal dashed line indicates the cut in $\delta(\rm{J}2000) = -2^{\rm{o}}$, below which almost no SDSS plate observations are available, and which increases the
    spectroscopic completeness.}
     \label{fig:radec}
\end{figure*}

Within a rectangular region tightly bounding the eFEDS X-ray coverage ($126^{\rm{o}}<\alpha(\rm{J}2000)<146.2^{\rm{o}}$, $-3.2^{\rm{o}}<\delta(\rm{J}2000)<+6.2^{\rm{o}}$, totalling 189.6\,deg$^2$), there are  61\,430 spectra listed in the DR16 catalogue, corresponding to 54\,743 unique astrophysical objects. 
These spectra are distributed fairly evenly over the eFEDS field, except for the South-West corner, which lies outside the footprint of the BOSS survey and is sparsely populated \citep[see Fig.~5 of ][]{Almeida2023_sdss_dr18}.
We consider only the best quality spectrum per object (\texttt{SPECPRIMARY}=1), and reject spectra having low signal-to-noise ratio\footnote{The S/N means the median SDSS signal-to-noise ratio over all valid pixels in the observed wavelength range 3280-9860\AA.} (S/N, in this case \texttt{SN\_MEDIAN\_ALL}$<$1) or unconstrained redshift uncertainties (\texttt{Z\_ERR}$\le$0).
This gives a sample of 49\,306 spec-z, of which 17\,906 were obtained with the original SDSS spectrograph, and 31\,400 with the BOSS spectrograph \citep{Smee2013}.
The majority (91\%) of these spectra were obtained as part of SDSS programmes carrying out galaxy and QSO redshift surveys (SDSS-Legacy, BOSS), and most of the remainder are associated with the SEGUE survey \citep{Yanny2009_SEGUE}.
A relatively small fraction of the objects associated with these spectra are expected to be detectable in the X-rays by a survey as deep as eFEDS \citep[see e.g.][for the cases of non-active galaxies and stars, respectively]{Vulic2022, Schneider2022}.
In contrast, the dedicated eFEDS plate observations, obtained as part of the SDSS-IV and SDSS-V surveys, have primarily targeted optical counterparts of eFEDS X-ray sources.
We refer the reader to \citet[][their Sect.~7.5]{Abdurrouf_2022_sdss_dr17} for a description of the SDSS-IV/eFEDS dataset (consisting of seven observed plates, with up to 1000 BOSS spectra obtained per plate). 
The dataset contains 6159 science spectra, although a significant minority of them (24\%) are of lower quality (\texttt{SN\_MEDIAN\_ALL}$<$1).
The vast majority of targets were selected as counterparts of eFEDS X-ray point-like sources and candidate clusters of galaxies. 

The SDSS-V/eFEDS target selection and dataset (comprising 37 plates) is described by \citet[][their Sects.~7.3 and 9.1]{Almeida2023_sdss_dr18}. 
This project, part of the Black Hole Mapper survey, was executed in the early phases of the SDSS-V survey, before the availability of the new robotic fibre positioner, which  enabled an automated transition from the traditional plate-based system \citep{Almeida2023_sdss_dr18}.
A number of plates were specifically designed, constrained by the amount of dark observing time available to the project, the desire to minimise the total number of drilled plug plates, and the overall capabilities of the plate system to place fibres on naturally clustered targets.
To reach the highest completeness possible, targets lacking existing high-quality spectroscopic observations (either from previous SDSS generations or other telescopes) were given a higher chance of receiving a fibre. 
During this phase, up to 500 BOSS fibres were available per plate (including 80 reserved for sky observations, and 20 fibres placed on spectrophotometric calibration stars).  
The eFEDS/SDSS-V dataset contains spectra for 13\,269 science targets, of which 12\,446 (94\%) are optical counterparts to eFEDS X-ray sources. 
Of these observed X-ray targets, roughly three-quarters are counterparts to point-like X-ray sources (i.e. AGN candidates), and the remainder are candidate members of X-ray or optically selected galaxy clusters.

We note that the target selection for the dedicated SDSS-IV/V plates in the eFEDS field was reliant on early reductions of the eROSITA/eFEDS X-ray dataset (version `\texttt{c940 V2T}'), and early attempts at cross-matching with optical-to-infrared counterparts provided by the DESI Legacy Survey \citep[DR8;][]{Dey2019}, SDSS DR13 \citep{Albareti2017_sdss_dr13}, and the Hyper Suprime-Cam Subaru Strategic Project \citep[HSC-SSP DR2;][]{Aihara2019}.
However, the differences between the X-ray catalogue (and counterparts) used for SDSS target selection, and that presented by \citep{Salvato2022} are modest.
For example, of the 9199 optical counterparts to point-like X-ray sources that were spectroscopically observed as part of the SDSS-V plate programme, 7584 (82\%) are still considered to be the `best' counterpart to an eROSITA Main sample source in the eROSITA EDR catalogue \citep{Brunner2022, Salvato2022}.
For the remaining 18\% of that sample, either their X-ray detection is no longer considered significant, or an alternative counterpart is now preferred.
The early (\texttt{c940 V2T}) X-ray catalogue and counterparts are not considered further in this work; we derive our science catalogue (see Sect.~\ref{sec:methods}), solely from X-ray sources that are included in the most recent (and reliable) eFEDS Main catalogue \citep{Brunner2022}, using the best optical-to-IR counterparts presented by \citet{Salvato2022}.

Figure~\ref{fig:radec} shows the layout of the eFEDS field, indicating the circular plates of SDSS-IV/V observations and the sources that belong to the SPIDERS program.  

\begin{figure}[t]
\centering
\includegraphics[width=\columnwidth]{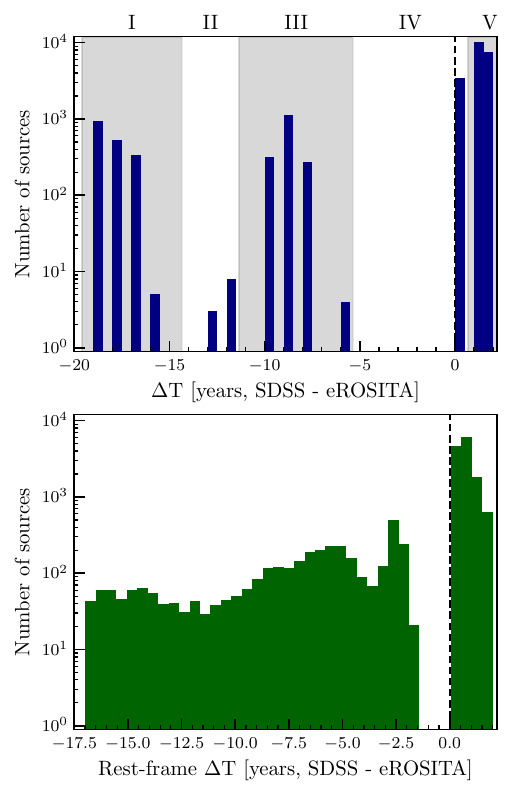}
  \caption{Time difference between observations from SDSS and eROSITA.
  The left panel shows MJD$_{\rm{SDSS}}-$MJD$_{\rm{eROSITA}}-$ in years.
  We highlight the different generations of SDSS as shaded regions, with SDSS I-III observations covering negative values (SDSS spectroscopy preceding eROSITA observations), while SDSS IV-V observations are to the right of the dashed line, indicating spectroscopic observations following the X-ray ones.
  The right panel displays the same time difference divided by $(1+\rm{z})$, showing the time difference in the rest-frame.
  }
     \label{fig:mjd_timediff}
\end{figure}

\begin{figure*}[t]
\centering
\includegraphics[width=\textwidth]{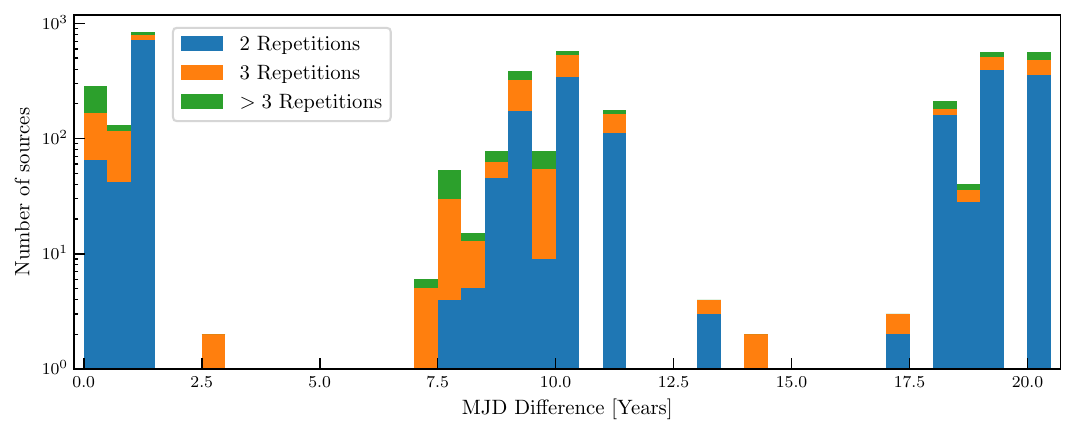}
  \caption{Difference in time between couples of SDSS spectroscopic observations of the same X-ray source, in years.
  The sources with 2 observations (blue), 3 observations (orange), and more than 3 observations (green) are summed up in this histogram display.
  }
     \label{fig:mjd_repetition}
\end{figure*}

Figure \ref{fig:mjd_timediff} shows the time difference between the observations of SDSS and eROSITA for each source, both in the observed frame (left panel) and in the rest-frame (right panel).
SDSS I-III observations that lie in the eFEDS field (negative values) will account for fewer objects and larger time spans, since these were not a follow-up of the eROSITA detections.
SDSS IV-V observations were taken with a maximum of one year and a half after the eROSITA observations and represent the majority of the observations described in this manuscript.
Having access to the information of when the data was taken can be interesting for looking for (correlated) variability, or for the study of changing-look AGNs \citep[see e.g.][]{Ricci2023}.
We also emphasise that some objects were observed multiple times by SDSS: $2\,451$ objects have two optical spectra, $341$ objects have three available spectra, and $73$ objects have more than three spectra.
These objects are displayed in Fig. \ref{fig:mjd_repetition}, where we show the difference in time between the repeated observations of the same object.
These multiple observations can be used in the search for variability when comparing the different SDSS spectra and looking for variations in their emission line fluxes and shapes, or also in their continuum shape.

\subsection{Non-SDSS spectroscopic redshifts in the eFEDS field}
\label{sec:additional_specz}

The eFEDS field has been observed by many spectroscopic surveys that are now public, most notably SDSS, GAMA, WiggleZ, 2SLAQ, LAMOST, and Gaia RVS, targeting a mixture of stars, galaxies, and QSOs.
The number of spectroscopic redshifts available within the eFEDS field aggregated across these surveys numbers in the hundreds of thousands.
Many of these spectroscopic redshifts are of high quality and can be used for science applications, in particular where one primarily only needs a redshift and a broad classification.
However, a careful collation and homogenisation of the existing spectroscopy catalogues is first needed to provide a reliable compendium of these data, as described below. 
The spectroscopic redshift catalogues used in this work are summarised in Table \ref{tab:specz_comp:inputs}.
In each case, we quote the number of selected spectroscopic redshifts that lie within a rectangular region tightly bounding the eFEDS X-ray coverage.
We have assigned each of these spectra a normalised redshift quality grade (\texttt{NORMQ}) according to the criteria described below. 
We expand further on the use and meaning of \texttt{NORMQ} in Sect.~\ref{sec:methods}. 

\begin{table*}[ht!]
\caption{Summary of catalogues and data sources providing spec-z for this project.}
\centering
\begin{tabular}{llcrrrrrr} 
 \hline
 \thead[l]{Parent\\Survey} & \thead{codename} & \thead{data \\ release} & \thead{$r_{\mathrm{match}}$ \\ (arcsec)} & \thead{Rank} & \thead{N$_{\mathrm{total}}$} & \thead{N$_{\mathrm{lead,comp}}$} &  \thead{N$_{\mathrm{lead,X-ray}}$} &  \thead{N$_{\mathrm{lead,X-ray}}$\\ \texttt{NORMQ}=3} \\ 
 \hline
SDSS    & sdssv\_vi  & dr18       & 1.0 &   1 &    13085 & 11739 &     6218 &     6011 \\
        & boss\_vi   & dr16       & 1.0 &   2 &     2843 &  2416 &     1014 &      992 \\
        & sdss\_vi   & dr16       & 1.5 &   3 &     2313 &  2239 &     1197 &     1187 \\
        & efeds\_vi  & dr17       & 1.0 &   4 &     6159 &  5855 &     2866 &     2237 \\
        & boss\_novi & dr16       & 1.0 &  10 &    31400 & 20674 &        0 &        0 \\
        & sdss\_novi & dr16       & 1.5 &  11 &    17906 & 10581 &        0 &        0 \\
GAMA    & gama       & dr4        & 1.5 &   5 &    74926 & 71237 &      239 &      212 \\
WiggleZ & wigglez    & final      & 1.5 &   6 &    20922 & 20146 &      159 &       75 \\
2SLAQ   & 2slaq      & v1.2       & 1.5 &   7 &      990 &   641 &       70 &       67 \\
6dFGS   & 6dFGS      & final      & 3.0 &   8 &      379 &   308 &       23 &       23 \\
2MRS    & 2mrs       & v2.4       & 5.0 &   9 &      152 &    71 &        7 &        7 \\
RCSED   & hectospec  & v2         & 1.5 &  12 &      352 &   193 &        2 &        0 \\
        & fast       & v2         & 3.0 &  13 &      375 &   115 &        0 &        0 \\
Gaia    & gaia\_rvs  & EDR3       & 0.5 &  14 &    15568 & 15571 &      699 &      699 \\
LAMOST  & LAMOST     & DR7v2.0    & 3.0 &  15 &    62997 & 53679 &      538 &      496 \\
SIMBAD  & simbad     & 2021.11.25 & 2.0 &  16 &    18094 &  2013 &       13 &        5 \\
NED     & ned        & 2021.11.25 & 2.0 &  17 &    51131 &  6585 &       34 &        0 \\
 \hline
 TOTAL & & & & & 334258 & 224063 & 13079 & 12011 \\
 \hline\hline
\end{tabular}
\tablefoot{A full description including references is provided in Sect.~\ref{sec:additional_specz}. 
$r_{\mathrm{match}}$ is the matching radius used when associating spec-z entries with the Legacy Survey DR9 catalogue.
Rank gives the priority order used when determining the `lead' catalogue per spec-z.
N$_{\mathrm{total}}$ gives the number of spec-z within the region bounded by $126^{\rm{o}}<\alpha(\rm{J}2000)<146.2^{\rm{o}}$, $-3.2^{\rm{o}}<\delta(\rm{J}2000)<+6.2^{\rm{o}}$ (encompassing the eFEDS X-ray footprint) which also satisfy the selection criteria listed in the text.
N$_{\mathrm{lead,comp}}$ gives the number of entries in the spec-z compilation where the (sub-)survey provides the `best' redshift per classification.
N$_{\mathrm{lead,X-ray}}$ gives the number of eFEDS main sample counterparts where the (sub-)survey provides the `best' redshift per classification, and the rightmost column gives the number of those which have high spectroscopic quality (\texttt{NORMQ}=3). 
The spec-z from SDSS are subdivided into several groups: 
sdssv\_vi and efeds\_vi - these are derived from the dedicated SDSS-V and SDSS-IV eFEDS plate programs, respectively, supported by visual inspections;     
boss\_vi and sdss\_vi - archival optical spectroscopy in the eFEDS field, obtained via the BOSS or original SDSS spectrographs, and supported by visual inspections;
boss\_novi and sdss\_novi -  archival BOSS and SDSS spectroscopy without visual inspections.
The data model for the spec-z compilation catalogue is described in Appendix \ref{sec:data_model_efeds_speccomp}.}
\label{tab:specz_comp:inputs}
\end{table*}

\noindent
\textit{GAMA}:
The location of the eFEDS field was chosen, in part, to overlap with the `GAMA09' sub-field of the Galaxy And Mass Assembly project \citep[GAMA,][]{Driver2009}.
The GAMA project obtained highly complete optical spectroscopy (spanning 3750--8850\,\AA) of (relatively) bright galaxies ($r<19.8$) to intermediate redshifts ($z\sim0.3$) obtained with the 2dF/AAOmega instrument \citep{Saunders2004} at the Anglo Australian Telescope. 
The GAMA survey covers over 60\,deg$^2$ of eFEDS \citep{Liske2015aa}, and so is expected to be particularly informative for both X-ray emitting galaxy clusters and the low-redshift, low-luminosity end of the AGN X-ray population.  
Here we use the catalogue of spectroscopic redshifts released as part of GAMA\,DR4 \citep{Driver2022_GAMA_DR4}, selecting 74\,926 spec-z entries having a spectroscopic quality grade (`NQ') of at least 2. 
We translate the GAMA quality grades into our normalised quality scheme (\texttt{NORMQ}) via $\texttt{NORMQ} = \mathrm{NQ} - 1$. 

\noindent
\textit{WiggleZ}: 
The WiggleZ survey \citep[][]{Drinkwater2010aa} primarily targeted UV-bright emission line galaxies at intermediate redshifts (0.2$<z<1$) as part of a cosmology redshift survey, also using 2dF/AAOmega. 
Approximately half of the eFEDS field is covered by the WiggleZ survey footprint.
We select 20\,922 spec-z entries from the final WiggleZ data release having a quality grade `Q' of at least 2.
We translate the WiggleZ quality grades into our normalised quality scheme (\texttt{NORMQ}) via $\texttt{NORMQ} = Q - 1$. 

\noindent
\textit{2SLAQ}: 
The 2SLAQ survey \citep[][]{Croom2009} targeted luminous red galaxies and optically selected QSO candidates with the 2dF instrument \citep[spanning 3700--7900\,\AA, ][]{Lewis2002}, and its footprint partially overlaps the eFEDS field.
We selected 990 spec-z entries from the 2SLAQ QSO catalogue \citep[v1.2, ][]{Croom2009} having quality grades (`qual2df') equal to 1 or 2 (and so ignoring 2SLAQ targets lacking a spec-z, and any spec-z that were derived from SDSS data).
We translate the highest 2SLAQ quality grade, qual2df = 1, to \texttt{NORMQ} = 3, and qual2df = 2 to \texttt{NORMQ} = 2. 

\noindent
\textit{6dFGS}: 
The Six Degree Field Galaxy Survey \citep[6dFGS, ][]{Jones2009} provides redshifts for a sample of well-resolved low redshift galaxies, derived from optical spectra (spanning at least 4000--7500\,\AA), which were obtained using the Six Degree Field multi-object spectrograph at the UK Schmidt Telescope.
The 6dFGS overlaps with the part of the eFEDS survey below $\delta=0^{\rm{o}}$. 
We select 379 spec-z from the final 6dFGS data release, having spectroscopic quality grade (`Q') in the range $3 \le \mathrm{Q} \le 6$. 
For 6dFGS spec-z with Q equal to 6 or 4 we assign \texttt{NORMQ} = 3, and \texttt{NORMQ} = 2 for the remainder.

\noindent
\textit{LAMOST}: 
The Large Sky Area Multi-Object Fiber Spectroscopic Telescope (LAMOST) survey has observed many bright objects (stars, galaxies, QSOs) within the eFEDS field.
We select spec-z derived from low-resolution LAMOST spectra (spanning 3690--9100\,\AA), as reported by the LAMOST DR7\_v2.0 data release \citep[][]{Luo2022}.
We apply the following selection criteria to reject low S/N and potentially problematic spec-z: S/N in the $r$ band (`snrr') of at least 5, non-null redshift, and a well-constrained redshift uncertainty (`z\_err'), resulting in 62\,997 spec-z.
For LAMOST spectra having $\mathrm{snnr} > 10$, $\mathrm{z\_err} < 0.002$ we assign \texttt{NORMQ} = 3, and assign \texttt{NORMQ} = 2 for the remainder.

\noindent
\textit{Gaia RVS}: 
The Gaia Radial Velocity Spectrograph \citep[RVS,][]{Sartoretti2018} has measured the radial velocities of bright stars over the full sky, covering a narrow wavelength range around the CaII IR triplet (8470--8710\,\AA). 
We select 15\,568 RVS measurements from Gaia EDR3 \citep{Gaia_EDR3},  all of which we assign \texttt{NORMQ} = 3.

\noindent
\textit{FAST} and \textit{Hectospec}: 
The Reference Catalog of Spectral Energy Distributions of galaxies project (RCSED) has compiled spectroscopic measurements for a large (4M objects) galaxy sample \citep{Chilingarian2017}.
We selected RCSED spec-z that had been derived from the FAst Spectrograph for the Tillinghast Telescope \citep[FAST, ][]{Mink2021}, and the Multi-Mirror Telescope Hectospec public archive \citep[][]{Fabricant2005}, reported as sub-samples of the `v2' RCSED database \citep{Chilingarian2024}.
We retain 375 FAST and 352 Hectospec spec-z within the eFEDS footprint, requiring snr\_median$ > 2$; all of which we assign \texttt{NORMQ} = 2.

\noindent
\textit{2MRS}: 
The 2MASS Redshift Survey obtained spectroscopy for bright NIR-selected galaxies over the whole sky, using various facilities. 
We selected 152 spec-z from the v2.4 release of 2MRS \citep[][]{Hucra2012}, all of which we assign \texttt{NORMQ} = 3.

\noindent
\textit{SIMBAD} and \textit{NED}: 
To collate additional archival spectroscopy from smaller-scale surveys and reported observations of individual objects in the eFEDS field, we have exploited spec-z information provided by the Simbad database \citep[][]{Wenger2000} and the NASA Extragalactic Database \citep[NED, ][]{Mazzarella2017}.
In each case, the archives were queried on 25 November 2021. To reduce duplication, we have attempted to filter out SIMBAD and NED entries that are associated with any of the spectroscopic surveys listed above.
We selected 18\,094 and 51\,131 spec-z entries from SIMBAD and NED, respectively. 
We translate Simbad quality metric grades A and B to \texttt{NORMQ} = 3.
We assign \texttt{NORMQ} = 2 for any lower quality Simbad spec-zs and for all spec-z originating from NED.

\section{Methods for compiling and validating the catalogue}
\label{sec:methods}

In this section, we describe the visual inspection procedure for the SDSS optical spectra, the collation of those visual inspections, and then finally how we have compiled the SDSS and non-SDSS redshift information into single redshift estimates per object in the eFEDS optical counterparts catalogues.  

\subsection{Visual inspection of the SDSS spectroscopy}
\label{sec:vi}

The BOSS redshift pipeline was originally designed and optimised for broad-line QSOs and passive red galaxies.
To verify the accuracy of such pipeline on the X-ray selected sources detected by eROSITA, we undertook an extensive visual inspection process of the entire collection of SDSS spectra of eFEDS point-like sources (14\,895 spectra of the 13\,079 sources followed-up with SDSS).
In this section, we briefly describe the process and summarise the main lessons learned. 
For a more detailed analysis of the quality and failure rate of the photometric redshifts in comparison to spec-z, we refer the reader to \citet{Salvato2022}.

The SDSS optical spectroscopic data reduction pipeline \citep[\texttt{idlspec2d},][]{Bolton2012} is used to fit a set of model spectral templates to each observed spectrum. 
The outputs include parameters that best fit redshift (and its uncertainty), object classification, and redshift warning flags.
For the vast majority of SDSS spectra that are not flagged by the pipeline as having uncertain redshifts, these pipeline-derived parameters are accurate \citep[as confirmed by inspection, or additional spectroscopic observations;][]{Bolton2012}. 
However, for a minority of spectra (as discussed below), the pipeline redshift estimates can lead to catastrophic failures.
Human visual inspection is required to identify and correct these failures. 
In addition, we can potentially use visual inspection to increase our confidence in lower-quality pipeline redshifts, thus increasing the size of the usable sample.

The visual inspection (VI) tools used by the SPIDERS team are described in \citet[][see their Sect.~3.5]{Dwelly2017}. 
In summary, a web tool is used to organise the efforts of a small team of volunteers (drawn from within the authors), to inspect a subset of spectra that require validation. 
The inspectors are presented with an interactive graphical visualisation of each spectrum overlaid with the best-fitting pipeline model (redshift+template).
The inspectors submit a response giving their opinion of the spectrum's true redshift and classification, as well as an encoded measure of their confidence in the redshift (see Table \ref{tab:vi_grading}).

We emphasise that we did not use artificial intelligence techniques while performing our visual inspection.
However, since the SDSS pipeline is based on machine learning, the results from the visual inspection (especially the corrections of the bad performances of the SDSS pipeline) can be later implemented in the training sample so the SDSS pipeline can be more reliable when dealing with the next and more numerous data releases.

\subsubsection{Details of the visual inspections and their collation}
The visual inspections are used to assign redshifts, classifications, and a simple normalised redshift quality metric (\texttt{NORMQ}) for each spectrum. 
Our definition of \texttt{NORMQ} is as follows: 3 - a secure spectroscopic redshift determined either via visual inspection or by a redshift fitting algorithm, 2 - a lower confidence spectroscopic redshift, e.g. derived from low S/N data or from a single emission line, 1 - a spectrum is available but it does not provide useful redshift constraints, -1 a spectrum visually determined to be Blazar-like, usually without strong redshift constraints, and finally, 0 - no spectroscopy is available.   

We have carried out two discrete rounds of visual inspections of SDSS spectra in the eFEDS field: phase I) inspection of SDSS-IV/eFEDS spectra obtained in March 2020 plus archival SDSS-DR16 spectra located near X-ray sources in the eFEDS field, and phase II) inspection of SDSS-V/eFEDS spectra obtained between Dec 2020 -- May 2021.  
For the pre-SDSS-V dataset (phase I) we carried out 16.3k inspections of 12.6k spectra, including 2.7k spectra examined by multiple inspectors. 
For the SDSS-V dataset (phase II) we performed 4.1k inspections of 3.6k unique spectra, focussing our inspection efforts primarily on spectra that lay in `higher risk' regions of the parameter space.
For example, we require complete coverage of the lowest and highest S/N tails of the population, spectra towards the extremes of the redshift range, and any spectrum with a pipeline redshift warning flag. 

An important step before exploiting the VI information was to collate and consolidate multiple inspections into a single estimate of redshift, classification, and confidence per spectrum.
This task was performed separately for each of phase I) and II). 
As a result of this collation step, a second round of re-inspection and homogenisation was done (by a small number of the most experienced inspectors) for a few hundred spectra that had received conflicting VIs or that had been flagged as worthy of further attention. 

The spectra were assigned collated \texttt{NORMQ} grades according to the hierarchical scheme given in Table \ref{tab:vi_grading}.
\begin{table*}
    \caption{Criteria used to grade the visual inspections of the SDSS spectra.}
    \centering
    \begin{tabular}{ccc}
    Criteria & \texttt{NORMQ} & notes \\
    \hline\hline
    \texttt{ZWARNING} bit 7 (=128) is set & 1 & An unplugged or dropped fibre\\
    \hline
    Noted as Blazar-like during VI  & -1 & A likely Blazar or BL~Lac object\\
    \hline
    ($S/N>2$ and $\mathtt{ZWARNING}=0$ and $0<z_{\rm err}<0.005$ & & \\
      and $N_{\rm conf3}\ge1$ and $z_{\mathrm{pipe}} < z_{\mathrm{max}}$ and $z_{\mathrm{pipe}} \sim \overline{z_{\mathrm{vi}}}$)  & & \\
    OR &  &  \\
    ($S/N>2$ and $\mathtt{ZWARNING}=0$& & \\
      and $N_{\rm conf3}\ge1$ and $stdev_i(z_{\mathrm{vi},i}/(1+\overline{z_{\mathrm{vi}}})) < 0.01$)  & 3 & Secure redshift and classification confirmed by VI \\
    OR &  &  \\
    ($S/N<2$ and $\overline{z_{conf}}>2.5$ & & \\
      and $N_{\rm conf3}>1$ and $stdev_i(z_{\mathrm{vi},i}/(1+\overline{z_{\mathrm{vi}}})) < 0.01$)  &  & \\
    \hline
    $\overline{z_{conf}}>2.0$ & & \\
    and $N_{vi}\ge1$ and $stdev_i(z_{\mathrm{vi},i}/(1+\overline{z_{\mathrm{vi}}})) < 0.01$  & 2 & Less certain redshift+classification provided by VI\\
    \hline
    $S/N>2$ and $\mathtt{ZWARNING}=0$ and $0<z_{\rm err}<0.005$ &  & \\
      and $N_{\mathrm{vi}}=0$ and $z_{\mathrm{pipe}} < z_{\mathrm{max}}$ & 3 & Trustworthy pipeline redshift+classification (no-VI)\\
    \hline
    Everything else & 1 & Low confidence in redshift+classification \\
    \hline\hline
    \end{tabular}
    \tablefoot{The criteria are tested in sequence (from top to bottom), so later grades must have failed the criteria for all previous grades. $N_{\rm conf3}$ is the number of visual inspections with \texttt{Z\_CONF}$=3$.
    The maximum trusted redshift depends on the classification, for CLASS=\texttt{QSO} we consider $z_{\mathrm{pipe}}<3$ safe and for CLASS=\texttt{GALAXY} we allow $z_{\mathrm{pipe}}<0.8$.
    The criterion $z_{\mathrm{pipe}} \sim \overline{z_{\mathrm{vi}}}$ is shorthand for $ | z_{\mathrm{pipe}} - \overline{z_{\mathrm{vi}}} | / (1 + \overline{z_{\mathrm{vi}}}) < 0.01$. 
    }
    \label{tab:vi_grading}
\end{table*}

\subsubsection{Results of collated VIs}
\label{sec:vi:results}

The main goals of the visual inspection of the objects are to guarantee that the redshifts attributed to the sources are reliable and to understand the possible failures of the SDSS pipeline to trace the potentially problematic cases in the next data releases.
Therefore, it is important to evaluate the performance of both the SDSS pipeline and its warning flag for unreliable redshifts (\texttt{ZWARNING} $\neq0$), and of the visual inspection procedure and its warning flag (\texttt{NORMQ} $<3$, see Table \ref{tab:vi_grading}).

\begin{figure}[!t]
\centering
\includegraphics[width=\columnwidth]{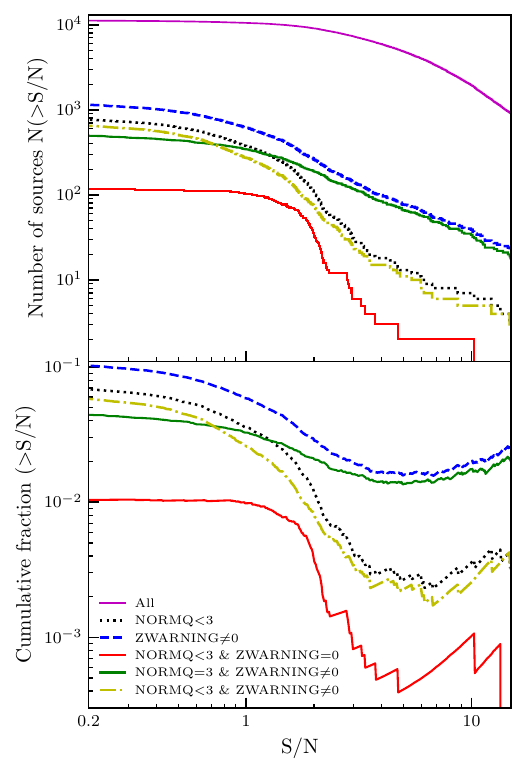}
  \caption{Diagnostic plot of the visual inspection.
  As a function of the median signal-to-noise ratio provided by the SDSS pipeline, we show the cumulative distribution (top) and cumulative normalised fraction (bottom) of different samples defined based on their SDSS pipeline (\texttt{ZWARNING}) vs. VI redshift measurement quality (\texttt{NORMQ}).
  The black dotted line represents the cases that are lost for science since their redshifts are not reliable according to the VI process ($\texttt{NORMQ}<3$).
  The blue dashed curve shows the cases for which SDSS data and/or pipeline fit are flagged as being problematic in some way($\texttt{ZWARNING}\neq 0$); the difference between the blue and the black lines shows the recovery of the VI.
  The red line indicates the most problematic cases, in which the visually inspected redshift is not reliable ($\texttt{NORMQ}<3$), but there was no indication from the SDSS pipeline that the redshift could be wrongly assessed ($\texttt{ZWARNING}=0$).
  The yellow dash-dotted line shows the other cases in which the redshift could not be recovered ($\texttt{NORMQ}<3$), but for which the SDSS pipeline indicated some problem ($\texttt{ZWARNING}\neq 0$).
  The combination of the red and yellow samples gives the total number of objects lost for science (black line).
  The green line represents the cases in which VI was needed ($\texttt{ZWARNING}\neq 0$) and successful ($\texttt{NORMQ}=3$).
 }
     \label{fig:SNRcum}
\end{figure}

Figure~\ref{fig:SNRcum} displays a performance check of the visual inspection compared to the cases where SDSS did or did not identify a failure in its redshift attribution.
The top panel shows the cumulative distribution of the median SDSS S/N (\texttt{SN\_MEDIAN\_ALL}), while the bottom panel exhibits the cumulative fraction\footnote{By cumulative fraction we mean the cumulative distribution of each sub-sample divided by the total number of objects above each given S/N threshold.} of spectra normalised by the number of spectra with the same S/N.
Here, the total sample is based on the full counterpart catalogue, with 27\,369 sources.
As expected, the total number of sources for each of the cases drops as the S/N increases, but the cumulative fraction increases from S/N$\gtrsim6$ since the increase of total sources is much less significant for higher S/N, making these populations more relevant.
Before the VI process (blue dashed curve),  $\sim10\%$ of all spectra have $\texttt{ZWARNING}\neq 0$, a fraction that drops to only $\sim$3\% for S/N$>$2 and stays around such proportion for higher S/N ($>$10).
The visual inspection allows the recovery of most failures at both high and low S/N, as shown by the green curve in comparison to the blue one.
Even after the visual inspection, at S/N $>10$ there are $\sim0.3\%$ cases in which the redshift is not obtained.
This result was expected from the X-ray selection, since the optical observation of some AGNs can present a typical power-law continuum, but without emission lines that allow tracing the redshift \citep[typically blazars; see also][]{Dwelly2017}.

In summary, we highlight the importance of the visual inspection to recover redshifts and provide confidence to the measurements for $\sim$99\% of S/N$>$2 spectra, and $\sim$ 94\% for S/N$>$0.2.
In general, this exercise also demonstrates that the BOSS pipeline is very efficient in deriving reliable redshifts for the X-ray selected sources, with a significant fraction of pipeline failures identified by the pipeline itself and corrected by our VI process (552 sources or $\sim2\%$ of the total sample, green line). 
The residual number of truly problematic cases (i.e., unreliable redshifts without any pipeline warning, red line) is small, plateauing at about 1\% for the whole sample and dropping to a sub-percent level for S/N$>$2 ($\sim0.4\%$ for the total sample, equivalent to 118 objects).

\begin{table}[t]
\caption{Classes attributed in the visual inspection of the eFEDS/SDSS spectra.}
\centering
\begin{tabular}{lcc} 
 \thead{VI Class} & \thead{Number of sources \\ (\texttt{NORMQ}$=$3)} & \thead{Percentage [\%] \\ (\texttt{NORMQ}$=$3)} \\ 
 \hline \\
\texttt{QSO} & 8\,524 (8\,456) & 75.5 (81.1) \\
\texttt{GALAXY} & 1\,622 (1\,599) & 14.4 (15.3) \\
\texttt{STAR} & 333 (331) & 2.9 (3.2) \\
\texttt{QSO\_BAL} & 41 (41) & 0.4 (0.4) \\
\texttt{BLAZAR} & 24 (0) & 0.2 (0.0) \\
\texttt{UNKNOWN} & 751 (0) & 6.6 (0.0) \\
\hline
Total & 11\,295 (10\,427) & 100 (100) \\
\end{tabular}
\tablefoot{The second column shows the total number of spectra and, in parentheses, the number of spectra that have a reliable redshift measured in the visual inspection process.
The third column lists the percentage of the spectra of each class with regard to the total SPIDERS sample and, in parenthesis, the sub-sample with reliable redshift.}
\label{tab:vi_class}
\end{table}

Another feature provided by the visual inspection is the attribution of classes to the objects.
Table \ref{tab:vi_class} presents the number of sources and percentages for the total sample and in parenthesis for the cases with reliable redshift (\texttt{NORMQ}$=$3).
The majority of the X-ray point-like sources are composed of AGNs with broad emission lines \citep[e.g.][]{Menzel2016}, for simplicity classified here just as \texttt{QSO}.
A small fraction of these objects also possess broad absorption lines, being classified as \texttt{QSO\_BAL} \citep[e.g.][]{Weymann1991, Rankine2020}.
We emphasise that all these objects with a clear absorption component in their broad emission lines have a high confidence in their estimated redshift.
The extragalactic sources with narrow lines in emission or absorption are all classified as \texttt{GALAXIES}.
Despite their X-ray point-like emission indicating the presence of an active supermassive black hole, if the AGN is obscured, the optical spectra can resemble star-forming galaxies, Type 2 AGNs \citep[hence the need of optical diagnostic diagrams such as in e.g.][]{bpt, veilleux_osterbrock1987, CidFernandes2011_WHAN, Mazzolari2024}, or passive galaxies \citep[][see also Sect.~\ref{sec:results:xray_pointlike}]{Fiore2003}.
We also note that, since the X-ray observations were not taken simultaneously with the optical follow-up, there could be intrinsic variabilities in the accretion phase of the observed AGNs.
All Galactic sources are classified as \texttt{STARS}, though they could be objects in different stages of stellar evolution, for example, cataclysmic variables, neutron stars, and white dwarfs \citep[see e.g.][]{Schneider2022, Schwope2024}.
These objects were identified in the visual inspection through stellar templates but mainly due to their continuum shape.
The objects classified as \texttt{BLAZARS} have a power-law continuum with no emission lines and, as mentioned before, do not have a reliable redshift.
Note that with this classification of \texttt{BLAZARS}, if a bona-fide blazar with prominent lines is observed, for example, an FSRQ \citep[see][and references therein]{Padovani2017}, it will have an associated redshift, and it will be classified as a \texttt{GALAXY}.
Also, weak-line quasars \citep[e.g.][]{Diamond-Stanic2009, Ni2018, Ni2022} could fall into this category.
Approximately 93\% of the sources classified as \texttt{UNKNOWN} have a noisy spectrum (S/N$<2$) that challenges a confident identification of their features.
The remaining cases exhibit only one emission line or an apparent overlap of sources, so the attributed redshift has low confidence.

\subsection{Building a compilation of all available spectroscopic redshifts in the eFEDS field}
\label{sec:spec_comp}

In this section, we describe how the wealth of spectroscopic information in the eFEDS field has been combined and homogenised into a single catalogue, providing a best estimate of classification, redshift and confidence per unique astrophysical object.

The eFEDS spec-z compilation presented here is intended to be as complete and versatile as possible, with only minimal (spatial and quality) down-selections from parent samples.
Therefore, this compilation includes many spec-z for non-X-ray sources, in addition to the many counterparts to eROSITA sources that are the main focus of this paper.

In cases where multiple data sources offer spec-z info for a single astrophysical object, we determine a single `best' spec-z.
We use the DESI Legacy Imaging Survey DR9 optical-to-IR catalogue \citep[][hereafter `LS9']{Dey2019} to define the list of all unique astrophysical objects within the eFEDS field.

The process of merging the multiple spectroscopic catalogues into a single spec-z measurement per astrophysical object is carried out by a dedicated Python package (\texttt{efeds\_speccomp}\footnote{\url{https://gitlab.mpcdf.mpg.de/tdwelly/efeds\_speccomp}}).
The algorithm for merging is as follows: 
i) Filter or down-select each input catalogue by the criteria described in Sect.~\ref{sec:additional_specz}, 
ii) normalise the heterogeneous spectral measurements (redshift, classifications, spectral quality) from each input catalogue onto a common system, 
iii) attempt to associate each input spec-z with any close LS9 objects (after removing \texttt{TYPE=DUP} entries, since they have no optical flux assigned to them), using a radial match\footnote{The matching radius per input catalogue used when compiling the spectroscopic compilation is given in Table \ref{tab:specz_comp:inputs} These radii were chosen to either match the fibre size, the typical object size, or were based on an empirical estimate of the consistency of the published coordinates w.r.t. LS9.} (after applying proper motions to move the LS9 objects to the approximate spectral epoch), 
iv) collate and rank (by standardised quality grade) all spec-z measurements for each unique LS9 object, then determine a final redshift, class, and normalised quality per LS9 object from amongst the top-ranked spec-zs for that object.

The last step (iv) can be broken into several steps, as follows.
For each unique LS9 object having at least one matching spec-z: 
a) collect the set of all available spec-z$_i$ for this LS9 object (subscript $i$ indexes over the available input spec-z), 
b) if at least one of the spec-z$_i$ has been visually inspected, then discard any that do not have a VI, 
c) discard any spec-z$_i$ that have a normalised quality grade lower than the maximum available normalised quality, 
d) set a flag if there is significant scatter amongst the redshifts of the remaining spec-z$_i$, 
e) set another flag if there is significant scatter amongst the normalised quality of the remaining spec-z$_i$
f) choose the `best' remaining spec-z$_{i,{\rm best}}$ according to the RANK of the input catalogue from which it was taken (see Table \ref{tab:specz_comp:inputs}), g) assign the \texttt{SPECZ\_REDSHIFT}, \texttt{SPECZ\_NORMC}, \texttt{SPECZ\_NORMQ} equal to those of the chosen specz$_{i,{\rm best}}$, set \texttt{SPECZ\_RANK} = 1 for these spec-z entries,
h) for the few spec-z which have no match to any LS9 object, we make no attempt to collate multiple measurements of the same astrophysical object, and they are propagated into the output catalogue without change.

The result of this process is a spec-z compilation catalogue containing a total of 334\,258 spec-z entries, corresponding to 222\,654 unique LS9 objects.
The catalogue also contains 1409 `orphan' spec-z which cannot be associated with any LS9 object.
The data model for the spec-z compilation catalogue is described in Appendix \ref{sec:data_model_efeds_speccomp}. 

\subsection{The eFEDS main source sample with updated spectroscopic redshifts}
\label{sec:efeds_main_cat_with_specz}

We matched the optical positions (taken from LS9) of the spectroscopic compilation catalogue to the optical positions of Main sample counterparts from \citet[][taken from LS8]{Salvato2022} using a conservative matching radius of 1\,arcsec.

We found spectroscopic redshifts for 13\,079/27\,369 ($\sim$48$\%$) of the counterparts from the eFEDS main sample, of which 12\,011 ($\sim$44$\%$ of the main sample and $\sim$92$\%$ of the assigned redshifts) have the highest spectroscopic quality rank (\texttt{NORMQ}=3). 
The number of objects with spectroscopic redshifts is approximately twice that presented in the earlier (pre-SDSS-V) work of \citep{Salvato2022}. 
Table \ref{tab:specz_comp:inputs} gives the number of `leading' spec-z provided by each contributing (sub-)survey for eFEDS main sample X-ray sources.
The objects with spec-z generally correspond to the optical brighter end of the sample, as we describe in greater detail below (Sect.~\ref{sec:results:general}).
The data format of this catalogue is described in Appendix~\ref{sec:data_model_efeds_main_with_speccomp}.

\section{Statistical properties of the spectroscopic sample}
\label{sec:results:general}

\subsection{Colour-colour diagram and X-ray to optical distribution}
\label{sec:color-color}

\begin{figure}[t]
\centering
\includegraphics[width=\columnwidth]{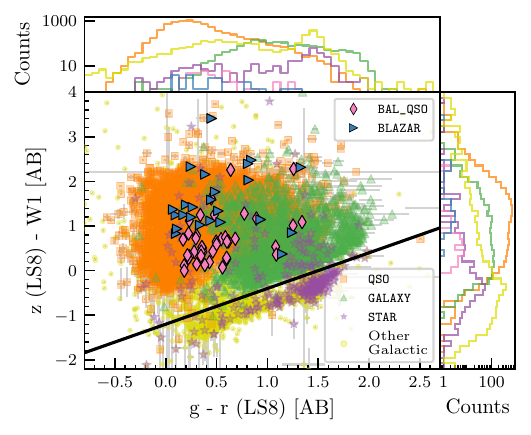}
  \caption{Colour-colour diagram of the sources with reliable redshifts (\texttt{NORMQ} $=3$), colour-coded by their class as in Table \ref{tab:vi_class}.
  For the colours $g-r$ and $z-W1$, $g$, $r$, and $z$ come from the Legacy Survey, while $W1$ comes from WISE.
  Orange squares represent broad emission line objects (\texttt{QSO}), while pink diamonds show the objects that have a broad line with an absorption feature (\texttt{QSO\_BAL}).
  Green triangles indicate extragalactic objects with narrow lines (\texttt{GALAXY}), and purple stars represent Galactic objects (\texttt{STAR}).
  The blue triangles represent \texttt{BLAZARS}, although these sources do not have reliable redshifts.
  Yellow circles indicate Other Galactic objects ($z<0.001$) that have a reliable counterpart, though they were not observed with SDSS.
  Both \texttt{BLAZARS} and Other Galactic objects do not have \texttt{NORMQ} $=3$.
  The black line divides objects between extragalactic (above) and Galactic (below) as in \cite{Salvato2022}.
  Histograms on the top and right show the distribution of each photometric colour per class.}
     \label{fig:grzw1_class}
\end{figure}

The spectroscopic classes defined in the previous section (see Table~\ref{tab:vi_class}) can also be visualised in a colour-colour diagram, as shown in Fig.~\ref{fig:grzw1_class}.
This colour-colour diagram with the filters \textit{grz} from Legacy Survey \citep{Dey2019} and \textit{W1} from WISE \citep{Toba2022} demonstrates the diversity of the X-ray selected sample, with Galactic objects below the black line and extragalactic objects above it, as proposed by \cite{Salvato2022}.
Although there is a tendency of the \texttt{GALAXIES} (which encompasses optical passive galaxies, star-forming galaxies, and Type 2 AGNs) to be located in the redder part of the diagram (right), and of the \texttt{QSOs} to be on the blue part of the diagram (left), the X-ray selection is effective at detecting broad-line emitters with red colours, complementing the QSO colour-selected in previous generations of SDSS as in \citealt{VandenBerk2001}.
This capability can be seen by the orange squares in the right part of Fig.~\ref{fig:grzw1_class}, showing an advantage in comparison to optical-only selections of AGNs \citep[as also seen in e.g.][]{Menzel2016}.
Regarding the rarer types of AGN that were classified with the visual inspection (\texttt{BAL\_QSO} and \texttt{BLAZARS}), they are mostly found in a region where typical quasars lie, but also with some redder objects.
We also plot Other Galactic sources, which are either stars or compact objects that have been detected as point-like sources with eROSITA and have a reliable counterpart but were not (yet) observed with SDSS.
Some of the Other Galactic objects lie in the extragalactic locus of the plot, but we do not have their spectra available to see if their redshift is properly assigned.

\begin{figure}[t]
\centering
\includegraphics[width=\columnwidth]{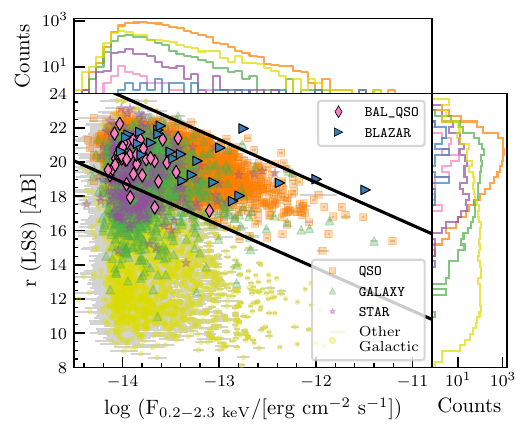}
  \caption{X-ray flux vs. optical magnitude diagram of the sources with reliable redshifts after the visual inspection, except for \texttt{BLAZARS} and Other Galactic sources which have \texttt{NORMQ} $\neq3$.
  The sources are colour-coded as in Fig.~\ref{fig:grzw1_class}.
  The region between the two lines defines the locus of quasars in this plane, adapted from \citet{Maccacaro1988}.}
    \label{fig:xray_r_class}
\end{figure}

Figure~\ref{fig:xray_r_class} displays the optical ($r$-band) magnitude vs. X-ray flux of the sample.
Investigation of possible trends comparing the X-ray emission and an optical filter dates back to at least \citet{Maccacaro1988}, with other large compilations found in e.g. \citet{Brusa2010}, or \citet{Civano2016}.
We adapted the equations that would define the locus of AGNs in the plane with the Legacy Survey DR8 $r$-band ($m_r$, central wavelength at $6420$ \AA\ and FWHM $1480$ \AA, see \citealt{Dey2019}) and the X-ray flux in the main eROSITA band ($F_X$, $0.2-2.3$ keV, considering the spectra to have a power-law index $\Gamma=2$ and a Galactic column density N$_{\rm{H}} = 3 \times 10^{20} \rm{cm}^{-3}$ as in \citealt{Liu2022}; see references therein for the motivation behind such values), with the flux ratios ($F_{\rm X}/F_{\rm O}$) corresponding to $10$ and $0.1$, respectively:
\begin{eqnarray}
    m_{r1} &=& -2.5\, \log (F_{\rm X}) - 16.19 \nonumber \\ 
    m_{r2} &=& -2.5\, \log (F_{\rm X}) - 11.19\,.
    \label{eq:maccacaro}
\end{eqnarray}
Hence, in comparison with the classes attributed via visual inspection (Table \ref{tab:vi_class}), the AGN locus defined by Eq.~\ref{eq:maccacaro} (Fig.~\ref{fig:xray_r_class}) is consistent with the distribution of \texttt{QSOs}, \texttt{BAL\_QSOs}, and \texttt{BLAZARs}.
Regarding the \texttt{GALAXIES}, $\sim$ 65$\%$ lie in the AGN locus, as expected from having a mixture of Type 2 AGNs, which should be in the AGN locus, but also star-forming and passive galaxies in the optical domain, which would lie below it.
Most of the objects classified as \texttt{STARS} with reliable redshifts are also inside the quasar locus ($\sim63\%$), but this result is also expected since this class encompasses cataclysmic variables, which are expected to have a relatively strong soft X-ray emission compared to stars and X-ray binaries.
For completeness, we also display the Galactic objects ($z<0.001$) with spectroscopic redshifts from other surveys than SDSS; they populate the region well below the AGN locus defined by Eq.~\ref{eq:maccacaro}.

\begin{figure}[t]
\centering
\includegraphics[width=\columnwidth]{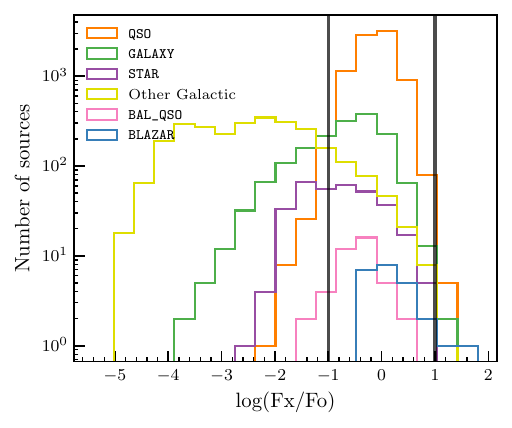}
  \caption{Distribution of the ratio of the X-ray flux ($F_{\rm X}$, flux at 0.2-2.3 keV from eROSITA) to the optical flux ($F_{\rm O}$, flux from the $r$-band magnitude from the Legacy Survey DR8).
  The classes are colour-coded as in Figs.~\ref{fig:grzw1_class} and \ref{fig:xray_r_class}.
  The vertical lines correspond to Eq. \ref{eq:maccacaro}.}
    \label{fig:hist_FxFo_class}
\end{figure}

A similar analysis can be done based on Fig.~\ref{fig:hist_FxFo_class}, where the ratio of the X-ray flux and the optical flux from the {\it r}-band are displayed.
The Other Galactic objects that were not observed with SDSS reach lower values of the X-ray flux in comparison to the optical flux, while the objects that have a reliable redshift after visual inspection have somewhat similar distributions of the $F_{\rm X}/F_{\rm O}$ ratio, with the exception of \texttt{GALAXIES} that show a larger tail to lower values of the ratio.

\subsection{Redshift distribution, completeness}

\begin{figure}[t]
\centering
\includegraphics[width=\columnwidth]{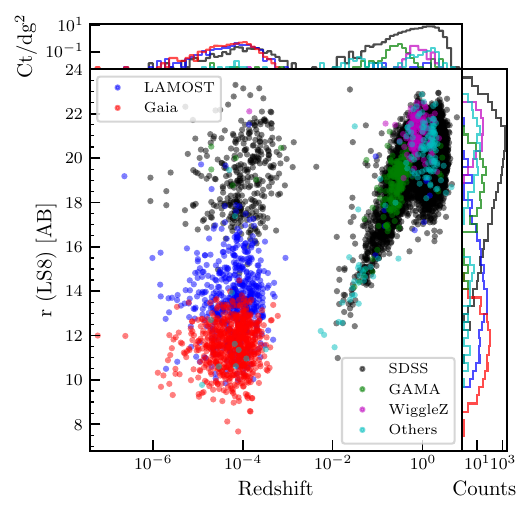}
  \caption{$r$-band magnitude as a function of the absolute value of the redshift, colour-coded by the survey that provided the redshift measurement (see N$_{\mathrm{lead,X-ray}}$ in Table \ref{tab:specz_comp:inputs}).
  The upper histogram shows the redshift distribution with regard to the number of counts divided by the eFEDS area ($140\ \rm{deg}^2$ as in the green contour of Fig.~\ref{fig:radec}, \citealt{Brunner2022}), while the histogram on the right displays the number of counts for $r$-band distribution.
  The absolute value of the redshift is plotted to allow the display of Galactic objects ($|z|\lesssim10^{-3}$) with negative radial velocities.
}
     \label{fig:r_redshift_surveys}
\end{figure}

The distribution of the spec-z compilation in both $r$-band magnitude and redshift is illustrated in Fig.~\ref{fig:r_redshift_surveys}, with the redshift spanning the range of $0\lesssim z \leq 5.8$.
We plot the absolute value of the redshift since Galactic objects could have negative values that correspond to their relative velocities more than the cosmic expansion, so their redshift does not necessarily represent their distance.
As seen in the upper histogram, there are two main groups of sources, with Galactic objects corresponding to $|z|\lesssim10^{-3}$ and the extragalactic objects with redshifts higher than this cut.
Among the Galactic objects, there is a wide spread in the observed $r$-band magnitudes, with SDSS providing most of the spec-z of fainter objects while surveys such as LAMOST and Gaia cover most of the bright sources.  
The main surveys that contribute to the redshifts of extragalactic objects in the eFEDS field are SDSS, GAMA, and WiggleZ; the cyan points represent all other surveys from Table \ref{tab:specz_comp:inputs} that did not have a significant number of sources when considered individually.

The quality of a sample can be determined by analysing its reliability and completeness.  
Reliability is the capability of selecting the aimed astrophysical objects without contamination of other types of objects in the sample, while completeness refers to the ability to select all the objects of such type in the sample without missing, for example, low-luminosity or highly obscured objects \citep{Hickox_Alexander_2018}.
In our case, the quantity of interest is the measurement of a spectroscopic redshift.

The visual inspections of the eFEDS spectra account for the reliability by assessing a quality flag to the determined redshift (\texttt{NORMQ}) and a class that indicates the nature of the observed spectrum (see Sect.~\ref{sec:vi}).
To address the completeness of the sample, we can compare the sample that has reliable spec-z with other subsets of the main eFEDS sample of the X-ray selected sources and their optical counterparts \citep[see][]{Salvato2022}.
From the main eFEDS X-ray sample (27\,369 sources), 13\,079 objects (48\%) have a spectroscopic redshift, with 14\,895 spectra.
Among those, 12\,011 objects (13\,674 spectra) have a reliable redshift according to visual inspection, corresponding to 44\% of the counterpart sample.

\begin{figure}
\centering
\includegraphics[width=\columnwidth]{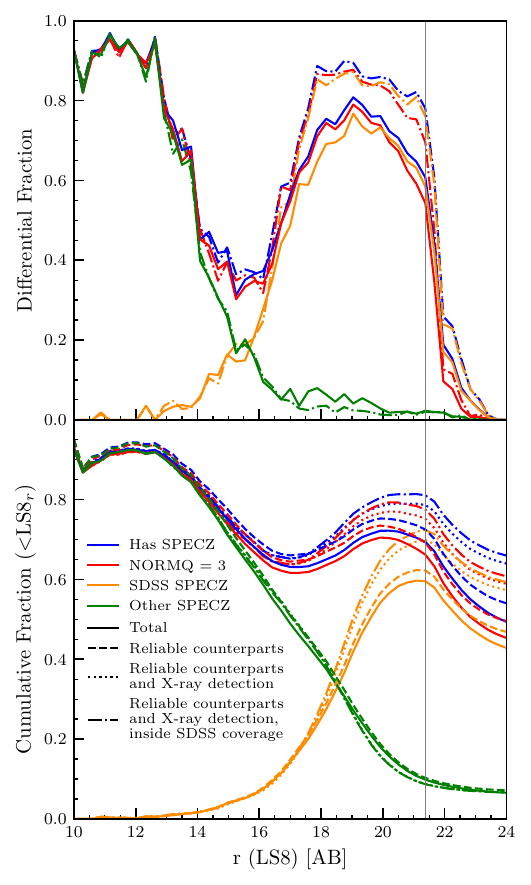}
  \caption{Differential (top) and cumulative (bottom) spectroscopic completeness fractions as a function of the r-band magnitude (top) or r-band magnitude limit (bottom).
  Blue lines are for all objects with a spectroscopic redshift from any survey (see Sect.~\ref{sec:spec_comp}), orange for those whose spec-z was obtained from SDSS, and green for other surveys.
  Red curves indicate the cases of high confidence for the redshift estimate after visual inspection.
  For each colour, the different parent samples are represented with different line styles: full sample (solid), only objects with reliable counterparts (\texttt{CTP\_QUALITY} $>1$; dashed), objects with reliable counterparts and a highly reliable X-ray detection (\texttt{DET\_LIKE} $>10$; dotted) and the most conservative case of objects with a reliable counterpart, reliable X-ray detection, and within the SDSS field coverage (\texttt{DEC} $>2$; dot-dashed).
  The vertical grey line indicates $r=21.38$, where the most complete sub-sample of objects with redshift from SDSS reaches its maximum fraction (dash-dotted orange line).}
     \label{fig:Rcum}
\end{figure}

Figure~\ref{fig:Rcum} assesses the completeness of sub-samples derived from the eFEDS catalogue.
The two panels show the distribution of the $r$-band magnitude from Legacy Survey DR8 (LS8), with the top panel showing the differential spectroscopic completeness as a function of magnitude, while the bottom panel displays the cumulative one as a function of the magnitude limit of the sample.
The peaks of the distributions can be explained by the limitations of the instruments used to obtain spec-z since sensitive spectrographs cannot observe too bright objects to avoid cross-talk effects on the spectra extracted on the CCD; this effect truncates the measured magnitudes and is more clearly visible in the upper panel. 
As expected, the sub-sample will be more complete as one cleans more of the comparison sample, indicating useful selections for dealing with the data.  
Comparing the different line styles, the total sample (filled lines) shows the lowest fractions in terms of completeness,
since it is more polluted and includes low-quality observations.  
The completeness is improved if we consider purer (smaller) samples satisfying other quality criteria, such as having a reliable counterpart (dashed lines, \texttt{CTP\_QUALITY} $>1$), a threshold on the X-ray detection likelihood (dotted lines, \texttt{DET\_LIKE} $>10$), and a cut in declination as indicated in Fig.~\ref{fig:radec} (dash-dotted lines, \texttt{DEC} $>2$).
The dip seen at $r\sim16-17$ between spec-z from SDSS and other surveys was seen also in Figure \ref{fig:r_redshift_surveys}, since BOSS design is aimed at observing fainter (and extragalactic) sources in comparison to other surveys that target mainly Galactic sources, such as LAMOST or Gaia.

\begin{figure*}
\centering
\includegraphics[width=0.9\textwidth]{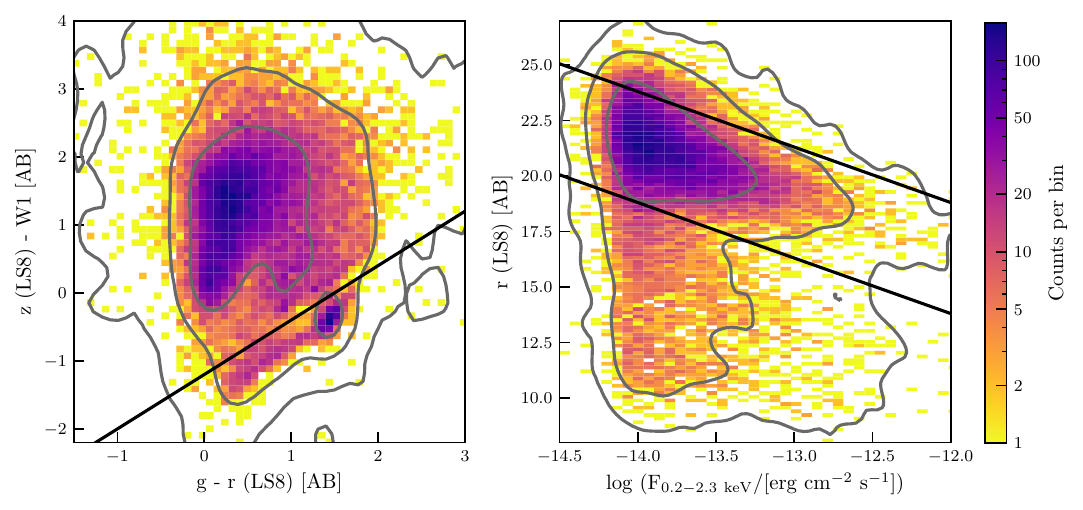}
  \caption{Colour-colour and flux-magnitude diagrams for the objects from the full counterpart sample with a reliable photometric counterpart (\texttt{CTP\_QUALITY} $>1$, 24\,774 sources), colour-coded by the data density per bin.
  The left panel presents $z$-W1 versus $g-r$, where $g$, $r$, and $z$ are obtained from Legacy Survey DR8, and W1 is from WISE Survey.
  The line separates extragalactic sources on the top from Galactic sources on the bottom, according to \cite{Salvato2022}.
  The right panel shows the $r$-band magnitude from the Legacy Survey DR8 versus the X-ray flux in the 0.2-2.3 keV band from eROSITA.
  The two lines adapted from \citet{Maccacaro1988} indicate the locus of quasars.
  Both panels exhibit their 1, 2, and 3$\sigma$ contours in grey.}
     \label{fig:density_ctpquality}
\end{figure*}

\begin{figure*}
\centering
\includegraphics[width=0.9\textwidth]{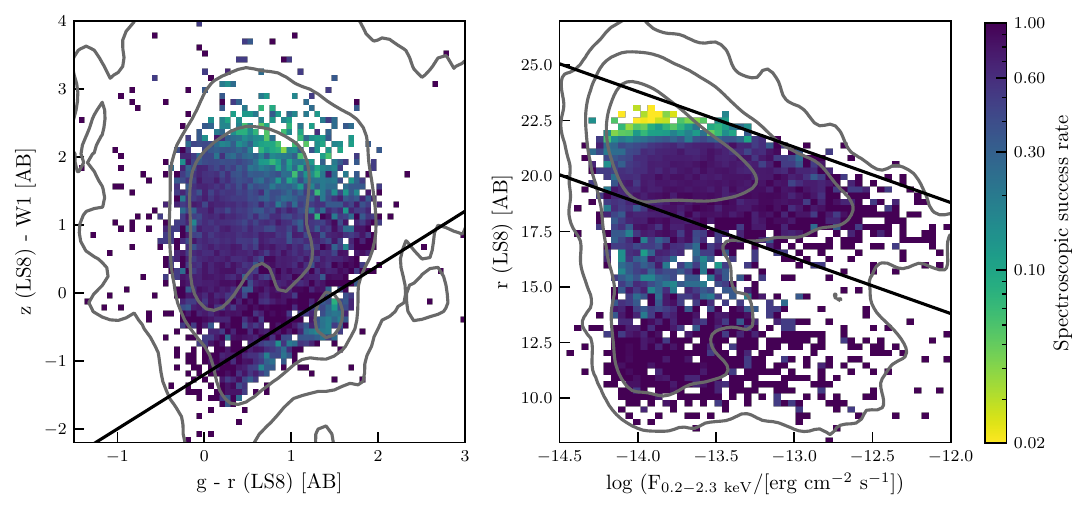}
  \caption{Colour-colour and flux-magnitude diagrams of the sources from the SDSS/eFEDS sample with reliable redshift (12\,011 sources) in comparison to those with a reliable photometric counterpart (24\,774 sources), as in Fig.~\ref{fig:density_ctpquality}.
  The colour-code is based on the spectroscopic success rate (i.e. N(\texttt{NORMQ} $=3$)/N(\texttt{CTP\_QUALITY} $>1$)).
  The grey contours indicate the same contours as in Fig.~\ref{fig:density_ctpquality} for the overall distribution of the X-ray sources with a reliable optical counterpart.}
     \label{fig:density_ratio}
\end{figure*}

Considering only objects with SDSS redshift (orange lines), for $r<21.38$ (below which the completeness drops rapidly), the total sample (filled line) will have spectroscopic completeness of 60\%, rising to 66\% for the sample with reliable counterparts (dashed line), to 70\% for the sample with reliable optical counterparts and high X-ray detection likelihood (dotted line), and increasing to 72\% for the sample that also accounts for the declination cut (dash-dotted line).
At the same magnitude threshold, the full sample with spectroscopic redshifts (including non-SDSS observations; blue) reaches 81\% of the contemplated objects, while the sample with reliable redshifts (red) reaches 77\%.

Figures~\ref{fig:density_ctpquality} and \ref{fig:density_ratio} present the spectroscopic completeness as a function of the location of sources in the colour-colour and flux-magnitude diagrams discussed above (Sect.~\ref{sec:color-color}).
Figure~\ref{fig:density_ctpquality} is colour-coded based on the density of sources with (\texttt{CTP\_QUALITY} $>1$) per bin, and the grey lines indicate such distribution in contours of 1, 2, and 3$\sigma$. 
The analysis of the completeness can be made based on Fig.~\ref{fig:density_ratio}, which is colour-coded by the spectroscopic success rate, i.e. the ratio of the number of sources with reliable redshift (\texttt{NORMQ} $=3$) to the number of sources with a reliable optical counterpart (\texttt{CTP\_QUALITY} $> $1) per bin.
The contours of Fig.~\ref{fig:density_ctpquality} are plotted in Fig.~\ref{fig:density_ratio} to facilitate the comparison.
It is clear from the top right of the left panel and the top left of the right panel that the SDSS observations of the sources with reliable counterparts do not cover the faint end of the distributions, due to instrumental limitations.  
The Galactic bulk of the left panel is less populated in Fig.~\ref{fig:density_ratio} when compared to Fig.~\ref{fig:density_ctpquality}, indicating that we are probably missing faint red stars, which are too faint for the Galactic programs used as LAMOST and Gaia (see Fig.~\ref{fig:r_redshift_surveys}).
Sources at $r\sim16$ have also lower spectroscopic completeness, but this is due to a combination of instrumental limitations and samples being observed (SDSS did not prioritise faint stars, while LAMOST and Gaia did) as exemplified in Fig.~\ref{fig:Rcum}.

\begin{figure}
\centering
\includegraphics[width=\columnwidth]{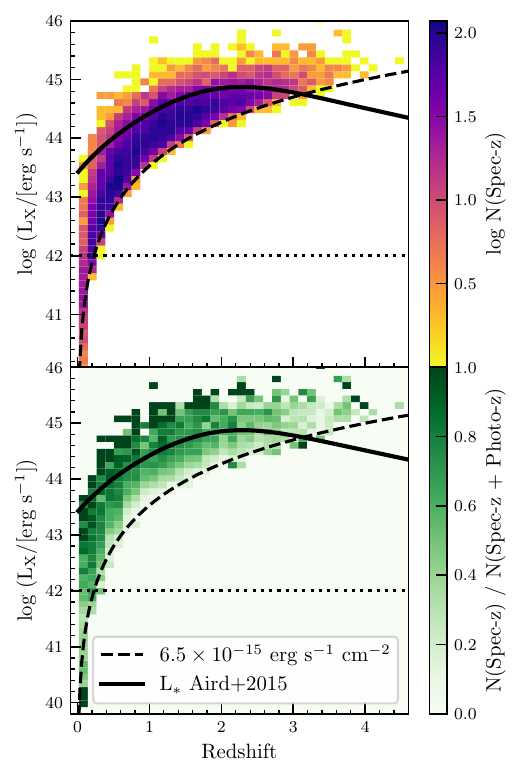}
  \caption{Observed X-ray ($0.2-2.3$ keV) luminosity distribution as a function of redshift.
  The top panel is colour-coded according to the number density in the logarithmic scale, while the bottom panel is colour-coded by spectroscopic completeness.
  The dashed line indicates the flux limit of the eROSITA survey, while the filled line indicates the location of the knee of the soft X-ray luminosity function according to the LADE model of \citet{Aird2015}.
  The dotted horizontal line indicates the limit of $L_{0.2-2.3 \rm{keV}}\leq10^{42}$ erg s$^{-1}$ applied in Sect.~\ref{sec:results:xray_pointlike} for a pure AGN selection.}     
    \label{fig:Xray_redshift_ratio}
\end{figure}
 
Figure~\ref{fig:Xray_redshift_ratio} shows the source distribution in the X-ray luminosity vs. redshift plane.
On the top panel we show the distribution for the objects with available spec-z.
On the bottom panel, for each of the 24\,774 objects with secure counterparts, we either assume the spectroscopic redshift, if available, or the photometric redshift as computed in \citet{Salvato2022}.
The plane is colour-coded by the spectroscopic completeness, i.e. the ratio between the number of sources with spec-z  and the total number of objects with either a photometric or spectroscopic redshift in the counterpart sample.
The X-ray luminosity was calculated from the eROSITA flux in the band between 0.2-2.3 keV (assuming a power-law spectrum with a spectral index $\Gamma=2.0$ and galactic absorption of $N_H=3\times10^{20}$ cm$^{-2}$ as described in \citealt{Brunner2022}) and the redshift, therefore not considering a K-correction.   
The dashed line indicates the approximate flux limit of eFEDS at $6.5 \times 10^{-15} \rm{erg\ s}^{-1} \rm{cm}^{-2}$ \citep{Brunner2022}, while the solid line indicates the knee of the soft X-ray AGN luminosity function ($L_*$) according to the LADE model from \citet{Aird2015}, after converting the Chandra band to the eROSITA band with the previous spectral index assumption.
We display the X-ray AGN luminosity function knee to show that our sample covers both its faint and bright ends up to a redshift significantly larger than one.

\section{SDSS spectral properties of extragalactic point-like sources: AGN stacks}
\label{sec:results:xray_pointlike}

In this section, we provide a global assessment of the physical and spectral properties of the AGNs in the SDSS/eFEDS sample.
We consider only objects that have reliable redshift (\texttt{NORMQ}$=3$) and optical counterpart (\texttt{CTP\_QUALITY}$>1$), and X-ray spectroscopic information from the \citet{Liu2022} catalogue. 
After such cuts, our sample consists of 10\,295 AGN candidates.

Among the advantages of the X-ray selection of AGNs is that it provides one of the most reliable AGN samples, since other astrophysical processes generating X-rays in a galaxy, such as X-ray binaries or hot gas, will be weak in comparison \citep[e.g.][]{Hickox_Alexander_2018}.
The majority of inactive galaxies have soft X-ray luminosities $L_{0.5-2.0 \rm{keV}}\leq10^{42}$ erg s$^{-1}$ \citep{Nandra2002, Menzel2016}.
A more detailed assessment of non-AGN contamination at lower luminosity necessitates reliable estimates of the hosts' stellar masses and star formation rates \citep{Lehmer2016, Igo2024}.
Lacking that information for all the objects in our sample, we adopted here a conservative cut of $L_{0.2-2.3\rm{keV}}>10^{42}$ erg s$^{-1}$ to guarantee that our AGN selection is as pure as possible.
This excluded 598 objects.

In addition to this cut, to account for a pure AGN selection, one should also account for misidentifications between the X-ray detection and the optical counterparts when associating the X-ray source with observations in other wavelengths \citep{Bulbul2022}. 
To guarantee that our sample is not selecting AGNs inside galaxy clusters, whose X-ray emission could be entangled with the emission of the hot gas within the cluster, we use only objects with \texttt{CLUSTER\_CLASS}$<4$ according to the optical counterpart matching described in \citet{Salvato2022}.
This excludes 42 sources.
Finally, we only consider sources with z\,$>0.001$ (based on Fig.~\ref{fig:r_redshift_surveys}) to guarantee that we are considering extragalactic sources, which excluded 301 objects from the total initial number of AGN candidates. 
All the above-mentioned cuts, which have some degree of overlap, result in a sub-sample with $9\,660$ objects discussed in this section (94\% of the total sample of AGN candidates).

Another advantage of AGN X-ray selection is the ability to detect both unobscured or obscured sources\footnote{Obscured AGN refers to objects with $N_H \geq 10^{22}\ \rm{cm}^{-2}$, while unobscured AGN refers to those with lower values of the column density.} \citep[e.g.][]{Ricci2022}, with obscuration depending not only on the orientation of the dusty torus but also on the host galaxy and large-scale environment where the AGN resides \citep[see e.g.][]{Brandt2015, Gilli2022, Andonie2024}.
As shown in \citet{Liu2022}, the eROSITA/eFEDS sample is biased towards unobscured AGNs due to the lower sensitivity of the pass-band beyond 2.3 keV ($\sim92\%$ of the sources with $N_H < 10^{21.5}\ \rm{cm}^{-2}$), which makes its Main Sample \citep{Brunner2022} focus on softer X-ray bands than other instruments.
The Hard Sample, presented in \citet{Waddell2024}, shows that eROSITA can detect obscured AGNs with column density up to $10^{24}$ cm$^{-2}$, though the Hard Sample contains much fewer objects than the Main Sample.
Hence, the objects detected at lower redshift do not have strong obscuration, while those at higher redshift have lower photon counts.
Another instrumental limitation to account for is the fact that both eROSITA and SDSS are flux-limited surveys, so faint sources that could be associated with low-luminosity AGNs should be underrepresented (see the discussion about Figs.~\ref{fig:density_ctpquality} and \ref{fig:density_ratio} for all the objects in the sample).

The X-ray selection overcomes some of the limitations of optical selection, especially for cases with obscuring material and faint or distant AGNs.    
Moreover, since the X-ray emission is less affected by the host-galaxy emission, it allows one to observe galaxies with different morphologies and in different stages of their evolution while still being able to separate the emission associated with the AGN from host galaxy emission.
Therefore, an X-ray-selected AGN sample will show a high degree of diversity in their optical properties, with passive galaxies, star-forming galaxies, LINERs, Seyferts (see e.g. \citealt{Heckman1980} and \citealt{Kewley2006} for the optical classification of local AGNs), and quasars.
Even in the case of quasars, the X-ray detections allow the selection of both blue and reddened quasars \citep[e.g.][]{Brusa2003, Menzel2016}. 

\subsection{Stacked spectral templates: Method}

\begin{figure}[t!]
     \includegraphics[width=\columnwidth]{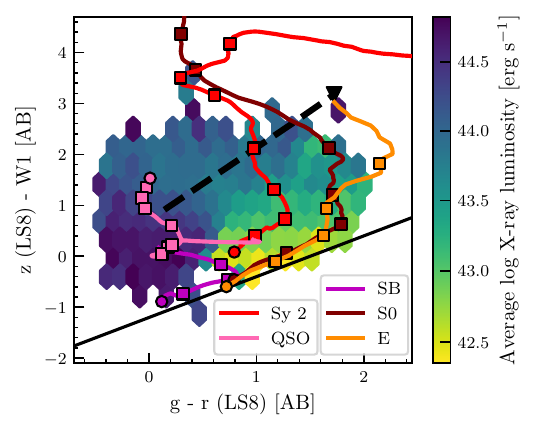}
     \caption{Colour-colour diagram with the AGN sample colour-coded by the average of the intrinsic X-ray luminosity (in the $0.2 - 2.3$ keV band) per bin, following \citet{Liu2022}. 
     The solid black line differentiates extragalactic objects on top from Galactic objects at the bottom according to \citet{Salvato2022}.
     The coloured tracks indicate the colour-colour evolution with redshift  of the SED templates from \citet{Salvato2022}, with $z=0$ shown as a circle.
     The purple track represents a spiral galaxy, brown represents a lenticular galaxy, orange represents an elliptical galaxy, red represents a Seyfert 2, and pink represents a quasar.
     The squares indicate the values of the redshift of $z=0.2,\ 0.4,\ 0.6,\ 1.0,\ 1.6,\ 2.0,\ 3.0$.
     The black dashed line indicates the trend of the quasar SED when more reddening is applied.}
\label{fig:results:grzw_xray}
\end{figure}

An example of the diversity of the AGNs in the eFEDS sample is shown in Fig.~\ref{fig:results:grzw_xray}, where we have the objects displayed over the whole locus of the extragalactic sources of the colour-colour diagram of \textit{gr-zW1} (as presented in Fig.~\ref{fig:grzw1_class}), colour-coded according to their intrinsic X-ray luminosity as measured by eROSITA \citep{Liu2022}.
Following the templates used in \citet{Salvato2022}, we show lines of different colours indicating tracks of different types of galaxies' SED between redshift of 0 and 3.
The black dashed arrow indicates the direction in which the quasar SED models are shifted after applying increasing reddening.
These tracks indicate the diversity of AGNs in our sample, and our objects can be described by at least one of such tracks when applying the correct redshift and reddening.  
Objects with lower X-ray luminosity populate the locus of (low-redshift) lenticular and elliptical galaxies; in contrast, the more intense X-ray emission is associated with bluer quasars, as expected.

\begin{figure}[t]
     \includegraphics[width=\columnwidth]{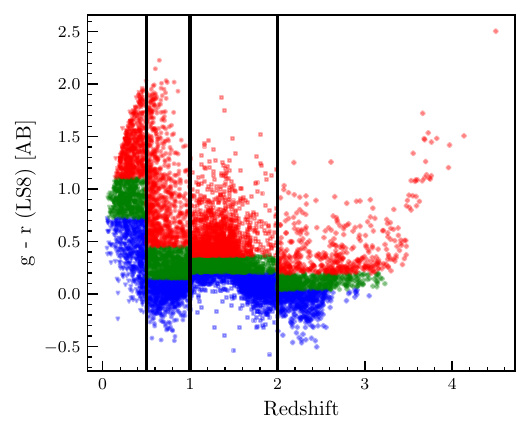}
     \caption{Colour ({\it g-r}) and redshift distribution of the data used in
       the stacks from Figs.~\ref{fig:results:stack_z05_full}-\ref{fig:results:stack_2z_zoom}.
       The stacks considered four redshift bins separated by vertical lines, with triangles at z $<0.5$, circles at $0.5\leq$ z $<1.0$, squares at $1.0\leq$ z $<2.0$, and diamonds at z $\geq 2.0$.
       Each redshift bin was divided into three colour ranges with approximately the same number of sources per colour bin, shown as blue, green, and red.}
     \label{fig:results:gr_redshift}
\end{figure}

\begin{figure}[t]
     \includegraphics[width=\columnwidth]{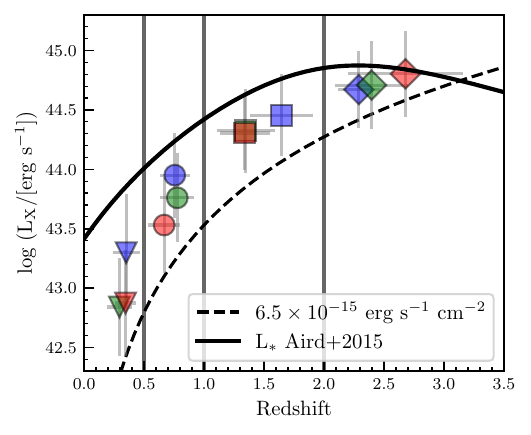}
     \caption{Mean intrinsic X-ray ($0.2-2.3$ keV) luminosity distribution as a function of mean redshift for the stacking bins of Fig.~\ref{fig:results:gr_redshift}, with the dashed and filled curves as in Fig.~\ref{fig:Xray_redshift_ratio}.}
     \label{fig:results:Xray_redshift_bins}
\end{figure}

After recognising the diversity of AGNs in our sample in colours and galaxy type, it is interesting to compare the spectral properties of these sources and describe the general features of the sample.
In this paper, we will only analyse the (stacked) optical spectra of the 9\,660 SDSS AGNs qualitatively, and the quantitative analysis with measurements of the features in each individual spectrum and their classification will be provided in a further paper (Aydar et al. in prep).
Since the stacks mix different classes of AGN, one should only consider trends such as comparing stellar populations and AGN power-law contributions to the observed continuum, the influence of the host galaxy, and line profiles, with caution to not overinterpret this averaged information.

The stacking technique consists of obtaining the median distribution of multiple spectra, and its main benefit is the increase of the S/N per pixel, which can unveil features in the spectra such as faint emission lines (see e.g. Sect.~5.1 of \citealt{Comparat2020}).
To define classes for the stacking by relying on information that does not depend directly on the spectral fitting of the sources, we considered the parameter space defined by redshift and g-r colour.  
Figure~\ref{fig:results:gr_redshift} shows the cuts we used to separate the objects into four redshift bins (low-z, $z<0.5$; mid-z, $0.5\leq z<1.0$; high-z, $1.0\leq z<2.0$; and very high-z, $z\geq2.0$) as vertical lines.
Then, for each redshift bin, we divided the sub-samples into three colour bins that have approximately the same number of objects per bin, with the intermediate colour bins for each redshift range as $0.72\leq(g-r)\leq1.1$; $0.15\leq(g-r)\leq0.45$; $0.21\leq(g-r)\leq0.37$; and $0.043\leq(g-r)\leq0.20$ for $z<0.5$; $0.5\leq z<1.0$; $1.0\leq z<2.0$; and $z\geq2.0$, respectively.
The similar number of objects stacked within each redshift bin provides spectra with a similar S/N, so larger uncertainties would indicate more diverse spectra being stacked and not just a statistical feature.  

To exemplify the groups of objects used for the stacks, in Fig.~\ref{fig:results:Xray_redshift_bins}, we show the average X-ray luminosity in the soft band of eROSITA \citep{Liu2022} as a function of the average redshift and respective standard deviations for each of the three colour bins in each of the four redshift bins.
The flux limitation of eROSITA and the soft X-ray luminosity function knee of \citet[][LADE model]{Aird2015} are also shown in Fig.~\ref{fig:Xray_redshift_ratio}.
We observe a trend in which the redder objects have a lower average X-ray luminosity than the bluer ones, which could be associated with obscuration.
This behaviour changes for the $z>2$ bin, though we can also note that the average redshift of the redder objects in such bin is higher than for the bluer objects, which could be indicating a selection effect of detecting the high-luminosity end of the red population.

\begin{figure*}[t]
     \includegraphics[width=0.9\textwidth]{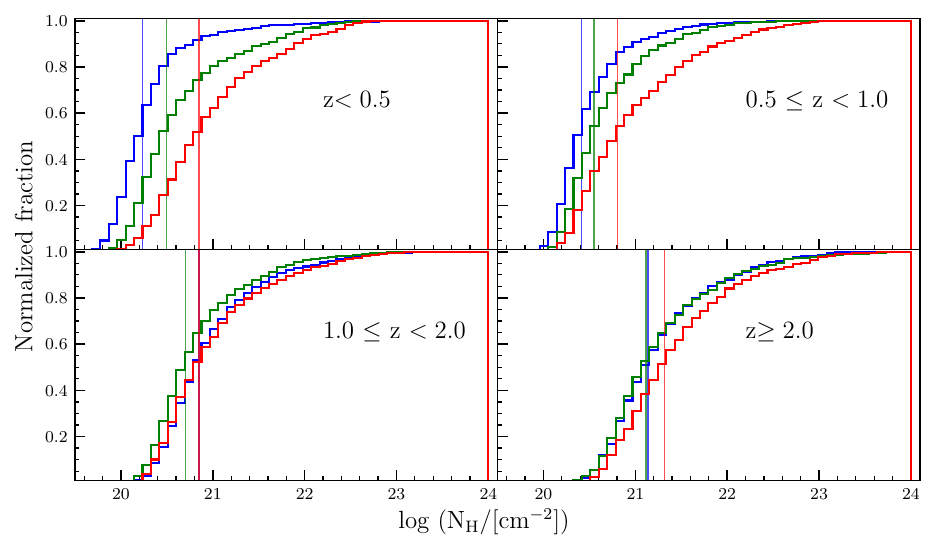}
     \caption{X-ray column density cumulative distribution per redshift and colour bin, as in Fig.~\ref{fig:results:gr_redshift}.
     The vertical lines indicate the median of the column density for each bin, with the colours corresponding to the colour bin (red for the redder bin, green for the intermediate colour and blue for the bluer objects).}
\label{fig:results:nH_colorbin_redshift}
\end{figure*}

With the X-ray information available, we can also analyse the behaviour of the X-ray absorbing column density\footnote{The column density values estimated by \citet{Liu2022} did not account for the more reliable redshift values published in this paper, so we are using the updated version of the eROSITA/eFEDs AGN catalogue available at \url{https://erosita.mpe.mpg.de/edr/eROSITAObservations/Catalogues/liuT/eFEDS_AGN_spec_V10.html}.} for each of the colour and redshift bins.
In Fig.~\ref{fig:results:nH_colorbin_redshift}, we see the trend of the low- and mid-z red bins having higher values of the median column density distribution, indicating that these objects are more obscured than the blue and intermediate colours.
For the high- and very high-z bins, however, the distributions are more similar, with the blue and red medians overlapping in the high-z panel, while the blue and intermediate colour medians overlap for the very high-z panel.
The overall column density is larger for higher redshifts, as expected, since the observed X-ray photons from further objects will be harder and, therefore, able to penetrate through thicker absorbers.

\subsection{Stacked spectral templates: Results}

\begin{figure*}[t]
     \includegraphics[width=\textwidth]{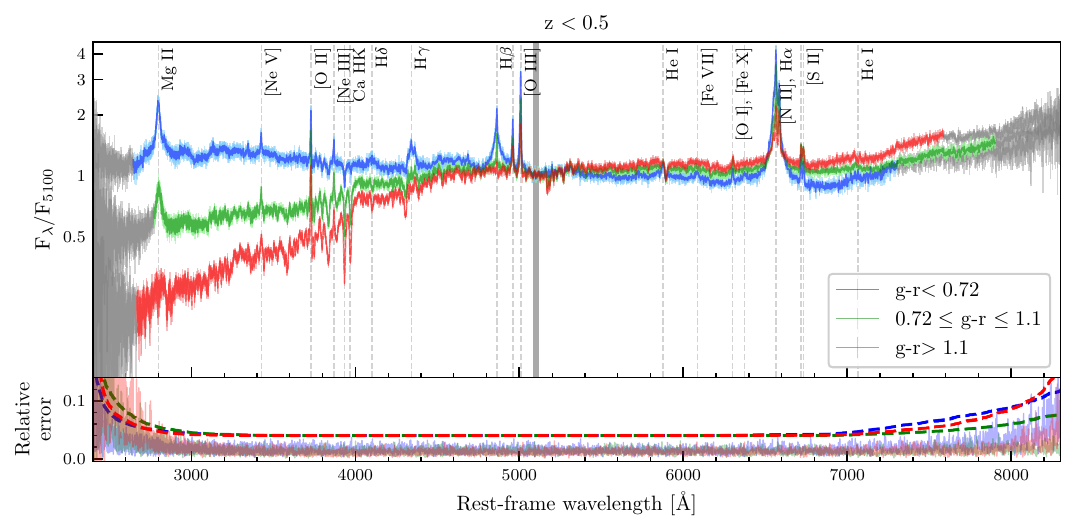}
     \caption{Median stacked spectra for AGNs with z $<0.5$, with the flux normalised at $5100$ \AA.
     The total sub-sample of z $<0.5$ AGNs was divided into three colour bins with 614 sources each.
     The top panel shows with coloured lines the wavelength coverage where more than $50\%$ of the total individual spectra contribute, with the remaining wavelength with less than half of the spectra in the stack shown in grey; the colour scheme is the same as Fig.~\ref{fig:results:gr_redshift}.
     Some of the main emission and absorption lines are indicated with dashed vertical lines.
     The grey-filled line indicates the wavelength range around 5100 \AA\ whose average flux was used for the normalisation.
     The bottom panel shows the errors calculated via the Jack Knife method in the stacking procedure divided by the median stack, and the dashed lines show $1/\sqrt{\rm{N}}$, where N is the number of individual sources per stacked pixel. }
\label{fig:results:stack_z05_full}
\end{figure*}

\begin{figure*}[t]
     \includegraphics[width=\textwidth]{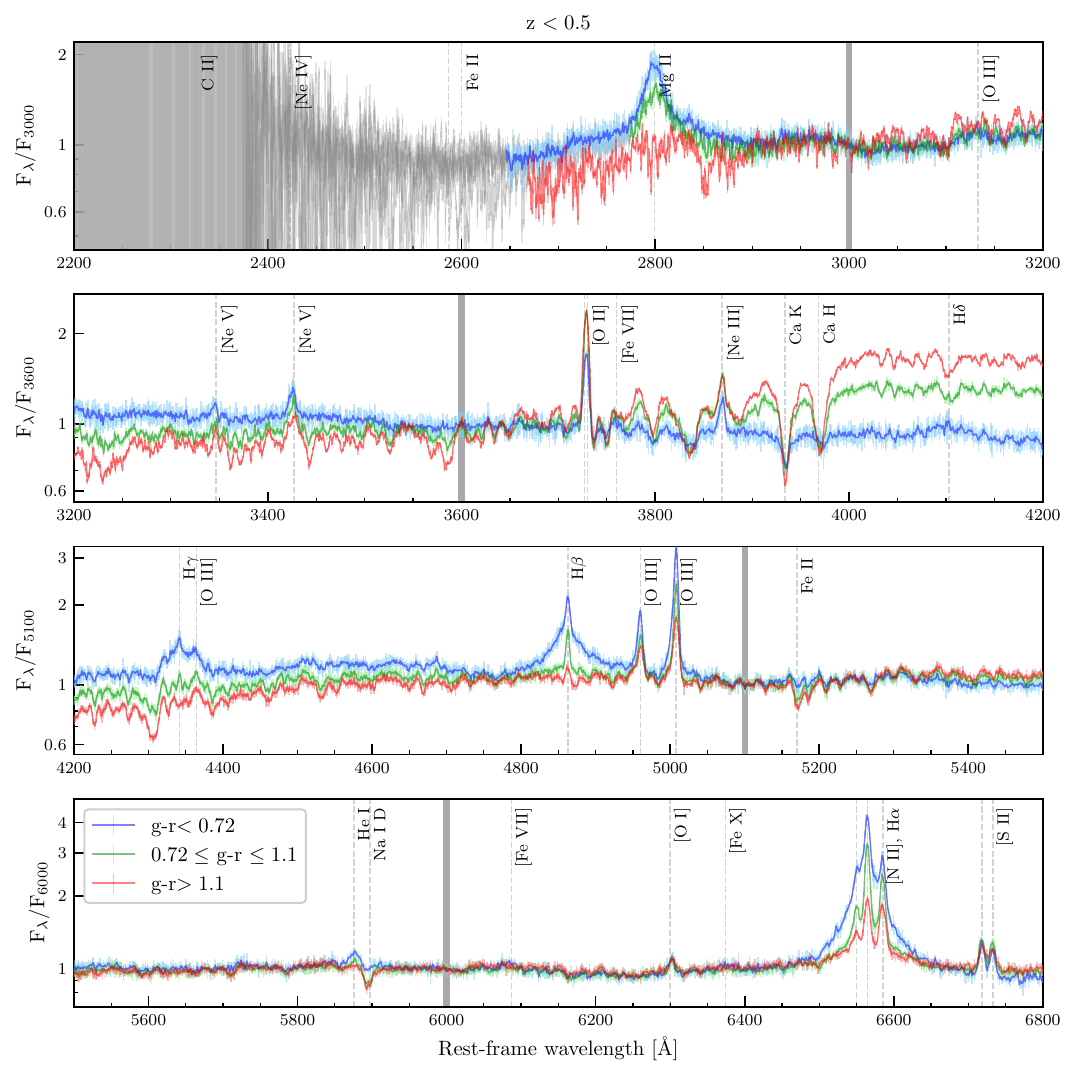}
     \caption{Zoomed-in version of the stacked spectra from Fig.~\ref{fig:results:stack_z05_full}. The emission and absorption lines are clearer, and the differences between the stacked colours are enhanced.
     The wavelength of the flux used for the normalisation of each plot is indicated in the ordinate axis and as a grey-filled vertical line.}
\label{fig:results:stack_z05_zoom}
\end{figure*}

\begin{figure*}[t]
     \includegraphics[width=\textwidth]{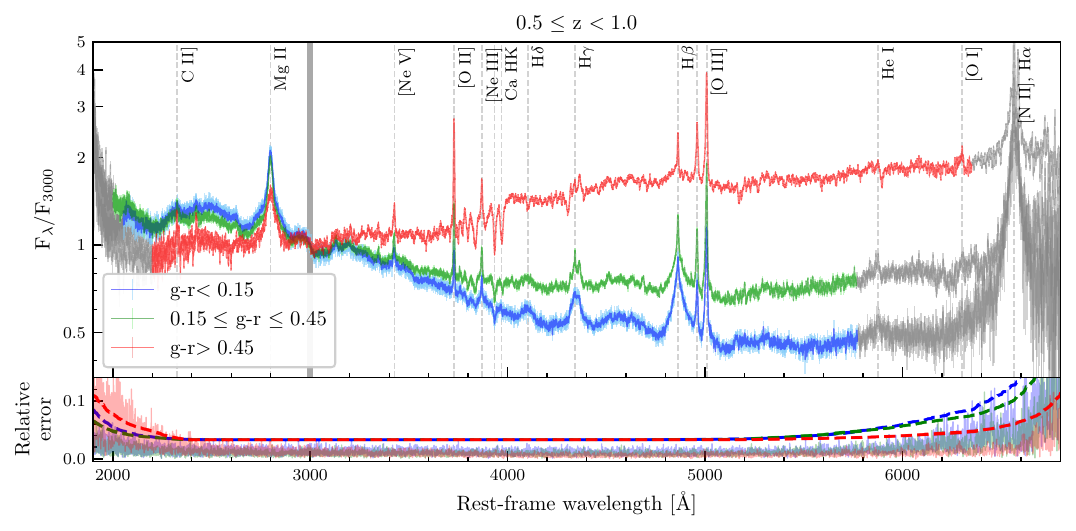}
     \caption{Similar to Fig.~\ref{fig:results:stack_z05_full}, but for $0.5\leq$ z $<1.0$ AGNs.
     The blue and red stacks contain 870 sources each, while the intermediate stack in green has 869.}
\label{fig:results:stack_05z1_full}
\end{figure*}

\begin{figure*}[t]
     \includegraphics[width=\textwidth]{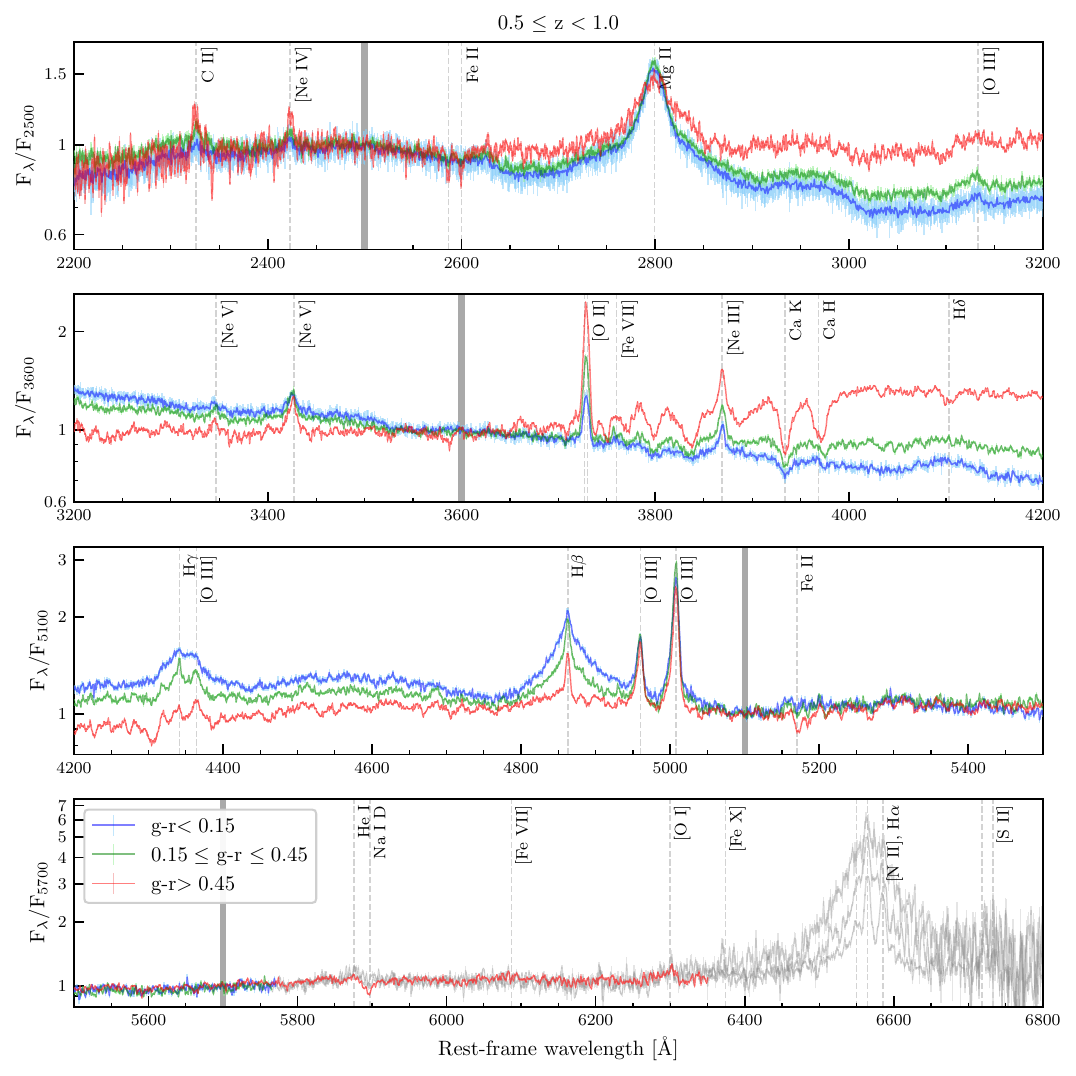}
     \caption{Zoomed-in version of the stacked spectra from Fig.~\ref{fig:results:stack_05z1_full}.
     The emission and absorption lines are clearer and the differences between the stacked colours are enhanced.
     The wavelength of the flux used for the normalisation of each plot is indicated in the ordinate axis and as a grey-filled vertical line.
     The wavelength range is the same as in Fig.~\ref{fig:results:stack_z05_zoom} to facilitate the comparison of the low-redshift stacks.
       }
\label{fig:results:stack_05z1_zoom}
\end{figure*}

Figure~\ref{fig:results:stack_z05_full} shows the median stacking of the low-z objects (z$<0.5$), with the upper panel showing the spectra of the blue, red and intermediate colours, and the lower panel showing the jackknife errors from the stack as filled lines and the number of objects per wavelength (as $1/\sqrt{\rm{N}}$) as dashed lines.
Figure~\ref{fig:results:stack_z05_zoom} shows a zoomed-in version of Fig.~\ref{fig:results:stack_z05_full} to allow focus on more details of each individual stack and the comparison between the different colours.
The other redshift intervals of $0.5\leq\rm{z}<1.0$ (mid-z), $1.0\leq\rm{z}<2.0$ (high-z), and z $\geq2.0$ (very high-z) are shown in Fig.~\ref{fig:results:stack_05z1_full}-\ref{fig:results:stack_05z1_zoom}, \ref{fig:results:stack_1z2_full}-\ref{fig:results:stack_1z2_zoom}, and \ref{fig:results:stack_2z_full}-\ref{fig:results:stack_2z_zoom}, respectively.
In each redshift sample, the spectra are normalised in a fixed wavelength interval with no prominent emission lines, shown as a grey-shaded region.
The main emission lines are highlighted with dashed vertical lines and the respective atomic transition is shown to the right of the line; the lines in Fig. \ref{fig:results:stack_z05_zoom} and \ref{fig:results:stack_05z1_zoom}, and in Fig. \ref{fig:results:stack_1z2_zoom} and \ref{fig:results:stack_2z_zoom} are the same, to allow a comparison of the low- and high-redshift cases.
We are showing in the colourful part of the spectra the wavelengths that contain more than 50\% of the total spectra included, while the grey extremities of the spectra show the wavelengths containing less than half of the total spectra in the stack.  

Out of the total AGN sample described in this section, one object is not considered in the stacks because it does not have an available g-band.
This source has the highest redshift of the SDSS/eFEDS catalogue, with $z=5.8$ after VI.

\subsubsection{Low-z AGNs (z$<0.5$)}
In the stacked spectra of the lowest redshift bin of z $<0.5$ from Fig.~\ref{fig:results:stack_z05_full} and \ref{fig:results:stack_z05_zoom}, we can see the presence of both emission and absorption lines.
Looking at the full spectra, normalised at 5100 \AA, we see that the red stack is more dominated by the host galaxy, with the blue stack being more quasar-dominated and the intermediate stack in green making a transition between such extremes.
The continuum of the red stack  has the shape typical of stellar populations, while the blue stack is flatter in the logarithmic scale, as expected for AGNs.  

To analyse the difference between the lines, looking at the zoomed-in plot is more instructive, where the normalisations are made in a part of the continuum closer to the wavelength of the lines.
Hence, we see in Fig.~\ref{fig:results:stack_z05_zoom} that the Mg II $\lambda2799$ emission, typical from quasar-dominated objects, despite being close to the limit of reliability of the spectra, is stronger in the blue than in the intermediate stack, and is not present for the red stack due to these spectra being more dominated by the host galaxy contribution than an unobscured broad-line AGN.
All stacks show [Ne V] $\lambda3426$ (and faint $\lambda3346$), though the coronal lines of [Fe VII] $\lambda3760$ and [Fe X] $\lambda6374$ are not distinguishable from the noise and, therefore, undetected.
The [O II] $\lambda3727,3729$ doublet and [Ne III] $\lambda3869$ line, often used in diagnostic diagrams to distinguish nuclear activity from star formation \citep{Kewley_Dopita_2002, Mazzolari2024}, are quite similar for the red and intermediate stacks, being weaker for the blue stack, as expected since the blue stack is more quasar-dominated and less affected by the emission from the stellar populations of the host galaxy as the red stack.
Confirming such a trend, the absorption lines of Ca KH and Na I D are similar for the red and intermediate stack (the red being more pronounced compared to the continuum than the intermediate for the Ca lines) while being weaker for the blue stack.
As for the Balmer line H$\delta$, it is in absorption for the red stack, in absorption with a faint emission peak for the intermediate stack, and within the noise for the blue stack, while the H$\gamma$ is in emission for the blue and intermediate stacks, but not identified in the red stack due to the more significant contribution of the host galaxy.

Regarding the BPT lines \citep{bpt, veilleux_osterbrock1987}, the [O I] $\lambda6300$ line and the [S II] $\lambda6717,6731$ doublet are similar for all the stacks.
The [O III] $\lambda4959,5007$ doublet has the same flux intensity pattern as the H$\alpha$ and [N II] $\lambda6584$ complex, being stronger for the blue stack and weaker for the red stack, with the intermediate colour between such extremes.
The H$\beta$ line, however, is stronger and broader in the blue stack, being less intense and narrower in the intermediate stack, and almost disappears in the red stack.
In the [O III] $\lambda5007$ line, it is possible to see the asymmetry of the line shape through the broader left wing, which could indicate the presence of outflows.

In addition, the Fe II pseudo-continuum emission is visible at both sides of H$\beta$ ($4440-4680$ and $5100-5400$ \AA) for the clue stack and becomes faint and mixed with the stellar continuum for the red stack, as pointed out for example by \citet{Negrete2018}, and quantified for instance by \citet{Bon2020}.

Therefore, all characteristics of the lines discussed above agree with the scenario in which the red stack is more dominated by stellar populations in the host galaxies, and the blue stack is more quasar-dominated, with the intermediate colours showing contributions from both quasar and star-formation activity.
It is also worth pointing out that the dispersion of the blue stack is significantly larger than the ones of the intermediate and red stack.
Since all stacks have the same number of spectra, this is explained by the fact that the blue stack contains a more diverse set of spectra.
This was confirmed with a visual inspection, where the main difference between the blue spectra is the power-law index of the continuum.
Also, the red stack contains more passive galaxies, which generally have more uniform and similar spectra.

\subsubsection{Mid-z AGNs ($0.5\leq\rm{z}<1.0$)}
A similar conclusion derived from the analysis of the z $<0.5$ stack can also be reached in the $0.5\leq\rm{z}<1.0$ stack shown in Fig.~\ref{fig:results:stack_05z1_full} and \ref{fig:results:stack_05z1_zoom}, where the red stack contains a more considerable contribution from the host galaxy while the blue stack is more AGN-dominated, with the intermediate colours in between.
However, in this case, the AGN component is clearly identified in all the stacks due to the strong Mg II emission.
At this intermediate redshift, we see the presence of typical AGN lines such as C II] $\lambda2326$ and [Ne IV] $\lambda2439$, and they appear more intense for the red stack due to the lower number of individual sources at this wavelength, which is also seen by the fact that the red stack seems noisier for shorter wavelengths.
Once again, it is possible to detect the [Ne V] doublet, and this time, the [Fe VII] $\lambda3760$ coronal line is also visible, though very faint compared to the noise, indicating that it is likely present only in a few sources.
The same trend of the red stack being more host-galaxy dominated is seen in the [O II] and [Ne III] emission lines and the Ca KH absorption lines.
The H$\delta$ line is in absorption for the red stack and within the noise for the intermediate and blue stacks.  

All of the BPT lines ([S II], [N II], H$\alpha$, [O I], [O III], and H$\beta$) are in the wavelength range with more than half of the total spectra, but the H$\beta$ shows the same trend as before of being broader and more intense for the blue stack compared to the red, with the intermediate stack as a transition.  
However, in this redshift range, the intermediate stack has the strongest [O III] $\lambda4959,5007$ emission, with the asymmetry of the line (a possible indicator of widespread outflows) being once again noticeable.

\begin{figure*}[t]
     \includegraphics[width=\textwidth]{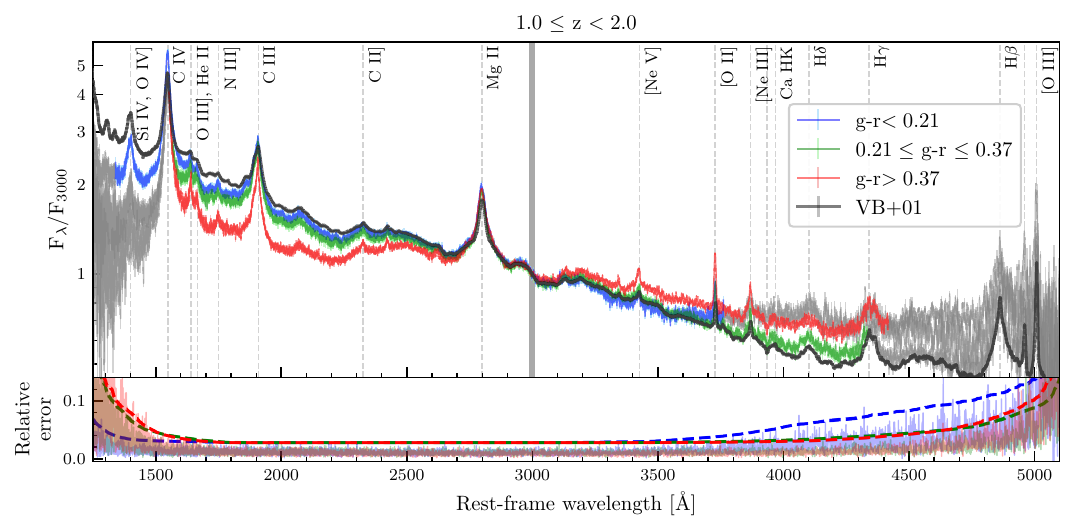}
     \caption{Similar to Fig~\ref{fig:results:stack_z05_full}, but for $1.0\leq$ z $<2.0$ AGNs normalised with regard to the flux at $3000$ \AA.
     The blue and red stacks contain 1271 sources each, while the intermediate stack in green has 1270.
     The black line indicates the QSO spectrum from \citet{VandenBerk2001}.}
\label{fig:results:stack_1z2_full}
\end{figure*}

\begin{figure*}[t]
     \includegraphics[width=\textwidth]{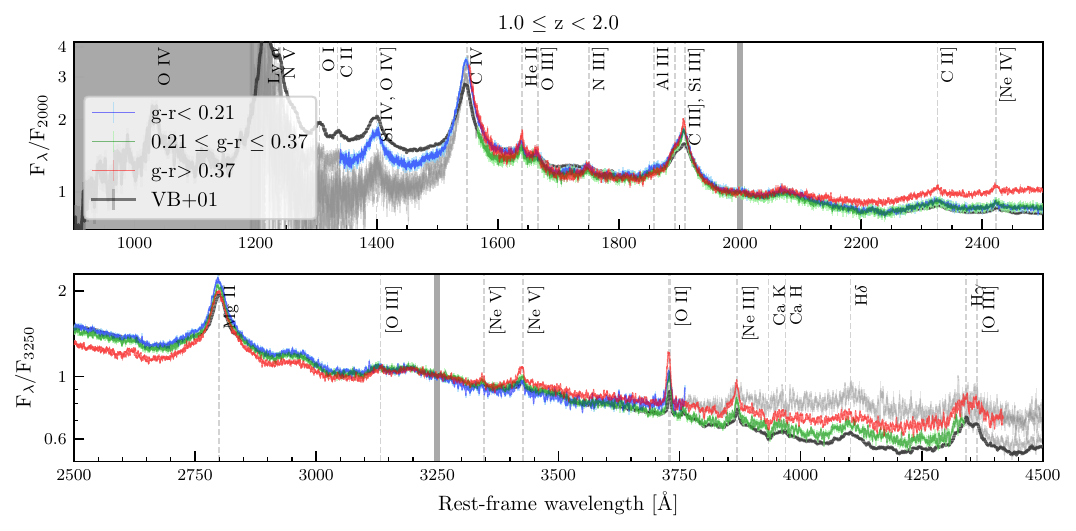}
     \caption{Zoomed-in version of the stacked spectra from Fig.~\ref{fig:results:stack_1z2_full}.
     The emission and absorption lines are clearer and the differences between the stacked colours are enhanced.
     The wavelength of the flux used for the normalisation of each plot is indicated in the ordinate axis.}
\label{fig:results:stack_1z2_zoom}
\end{figure*}

\subsubsection{High-z AGNs ($1\leq\rm{z}<2$)}
Figures~\ref{fig:results:stack_1z2_full} and \ref{fig:results:stack_1z2_zoom} show the stacked spectra for $1.0\leq\rm{z}<2.0$, where the most prominent optical emission lines used in AGN diagnostic diagrams fall outside the wavelength coverage, but the broad emission lines typical from quasars are present.
At those redshifts (and luminosities), the spectra are almost all QSO-dominated, as seen from the broad emission lines.
For comparison, we also plot the quasar template obtained from the stack of observed spectra of SDSS 1 QSOs from \citet{VandenBerk2001}, selected based on the optical SDSS filter system.
The quasar template mainly overlaps with the blue stack but is bluer at the edges of the covered wavelength range, indirectly suggesting the efficiency of the X-ray selection in detecting reddened quasars.

Regarding the Fe II pseudo-continuum emission, since we do not detect faint objects (that could include host-galaxy dominated AGNs), the optical Fe II becomes stronger as seen in the Vanden Berk spectra of Fig.~\ref{fig:results:stack_1z2_full}.

\begin{figure*}[t]
     \includegraphics[width=\textwidth]{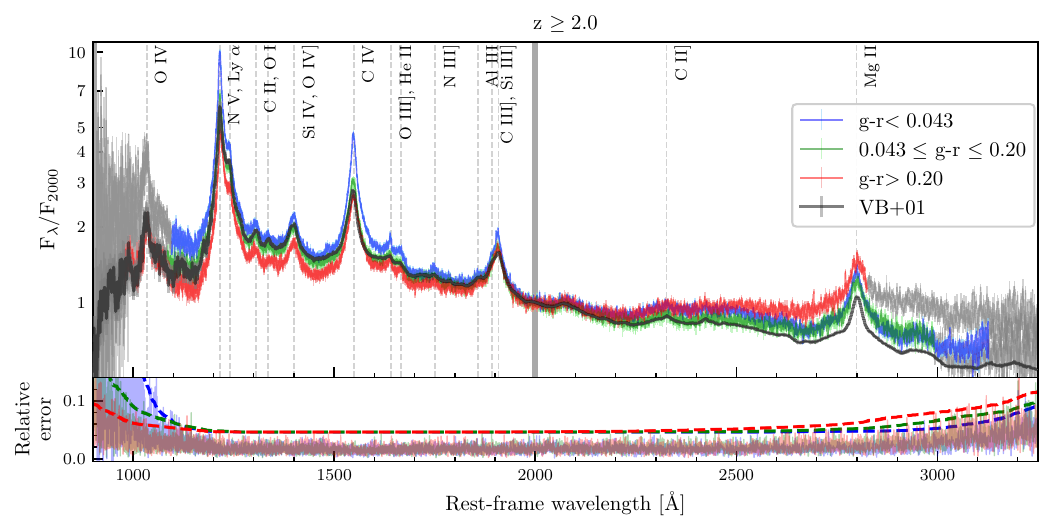}
     \caption{Similar to Fig~\ref{fig:results:stack_z05_full}, but for z $\geq 2.0$ AGNs normalised with regard to the flux at $2000$ \AA.
     The green and red stacks contain 465 sources each, while the blue stack has 466.
     The black line indicates the QSO spectrum from \citet{VandenBerk2001}.}
\label{fig:results:stack_2z_full}
\end{figure*}

On the zoomed-in version of the stack, some emission lines typical from quasars are seen, as C III] $\lambda1909$, Si III] $\lambda1892$, faint Al III $\lambda1857$, N III] $\lambda1750$, O III] $\lambda1665$, He II $\lambda1640$, and C IV $\lambda1549$.
As expected from quasars, they all have similar shapes and intensities when the continuum is normalised close to their wavelength.
The other lines that are typically seen on quasars, such as CII], [Ne IV], and Mg II, are also very similar, as is the typical AGN line of [Ne V].
Despite the difference in the continuum, which was already expected from the colour binning, the other difference between the stacks appears in the [O II] and [Ne III] emission lines, which trace star formation, being stronger for the red stack, consistent with the lower redshift bins.
The red stack shows broad Mg II, but narrow H$\beta$ (with a possible broad base).
This should be attributed to the fact that in such different wavelength ranges, the type of AGNs dominating the stack should be different, with more AGN-dominated objects in the higher redshift and lower wavelengths in comparison to the host galaxy-dominated objects in the lower redshift and higher wavelength.
This trend is also clear when comparing the red stacks at lower redshift (Fig.~\ref{fig:results:stack_z05_full}) to the ones at higher redshift (Fig.~\ref{fig:results:stack_1z2_full} and \ref{fig:results:stack_2z_full}).

\begin{figure*}[t]
     \includegraphics[width=\textwidth]{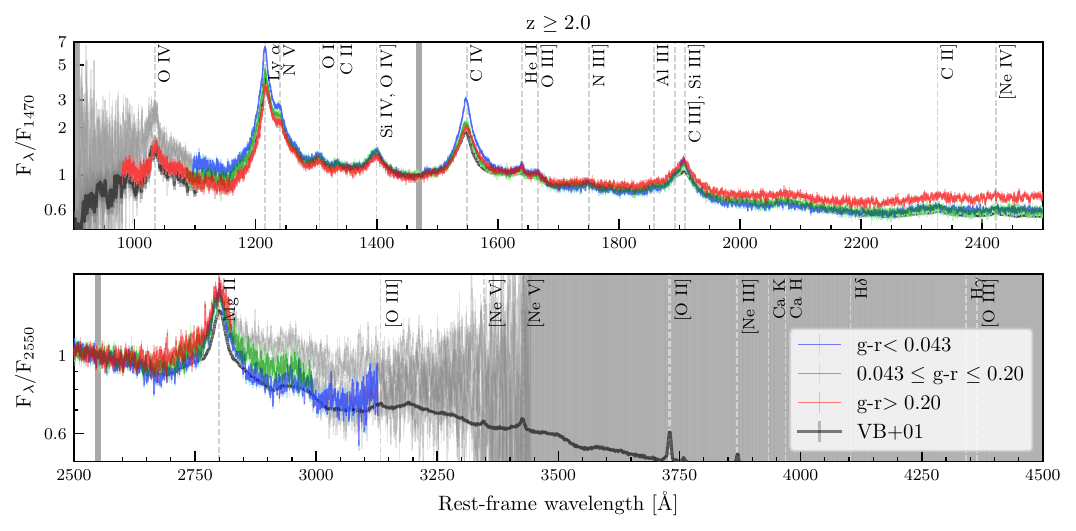}
     \caption{Zoomed-in version of the stacked spectra from Fig.~\ref{fig:results:stack_2z_full}.
     The emission and absorption lines are clearer and the differences between the stacked colours are enhanced.
     The wavelength of the flux used for the normalisation of each plot is indicated in the ordinate axis.
     The wavelength range is the same as in Fig.~\ref{fig:results:stack_1z2_zoom} to facilitate the comparison of the high-redshift stacks.}
\label{fig:results:stack_2z_zoom}
\end{figure*}

\subsubsection{Very high-z AGNs ($\rm{z}>2$)}
Finally, Figs.~\ref{fig:results:stack_2z_full} and
\ref{fig:results:stack_2z_zoom} show the stacked spectra of the z$\geq2.0$ bin, also together with the quasar template from \citet{VandenBerk2001}.
Except for the larger wavelengths, once again, the quasar template overlaps the blue stack, as expected, given the flux limit of eFEDS and the prevalence of unobscured AGNs at the highest luminosities.
More lines can be identified, such as the blend of Si IV and O IV] at $\lambda1400$, C II $\lambda1335$, O I $\lambda1305$, N V $\lambda1240$, Ly$\alpha$, and O IV $\lambda1033,1038$.

\section{Outlook and conclusions}
\label{sec:conclusions}

We reported on one of the largest systematic spectroscopic follow-up studies in X-ray survey fields, that was driven by the SDSS follow-up of eROSITA sources in the eFEDS field.
The X-ray spectra cannot be interpreted without the redshift of each object.
We therefore compiled all available redshifts for the eROSITA/eFEDS sources and described the compilation in Sect.~\ref{sec:data}.
We considered surveys including SDSS, GAMA, WiggleZ, Gaia, and LAMOST (see Table \ref{tab:specz_comp:inputs} for a summary).
Out of the total 27\,369 point-like X-ray sources detected with eROSITA in the eFEDS field, 13\,079 objects have an associated spectroscopic redshift.
Some objects were observed more than once, and this resulted in 14\,895 total spectra of stars, galaxies, and QSOs. 

With the initial estimates of the redshifts obtained from archival public data, the SDSS spectroscopic follow-up allowed us to confirm the reliability of the redshifts and of the SDSS pipeline through a meticulous visual inspection process that we described in Sect.~\ref{sec:methods}.
Using strict quality cuts that were mainly informed by this inspection process, we confirmed reliable redshift estimates for 12\,011 objects.
We were able to reliably recover the redshifts for 99\% of the sample with S/N$>2$.
The remaining cases are either spectra without clear emission or absorption lines or objects overlapping within the fibre.
The reliable assessment of the redshift was more difficult for noisier spectra, but we were still able to recover 94\% of the sample with S/N$>0.2$.
The visual selection process also allowed us to assign a (coarse) class for each spectrum and to further study the behaviour of the populations in different parameter spaces.
This showed us that, as expected, our sample of point-like X-ray sources is dominated by AGNs.
Only $\sim3\%$ of the SDSS spectroscopic sample consists of Galactic objects.
The goal of this VI approach was to use human power to understand potential failures and improvements of the SDSS pipeline.
This set of data might be very useful for training machine-learning techniques for the larger amount of data that will be available in the next data releases.

After flagging the reliable redshifts, we discussed the completeness and purity of the spectroscopic catalogue in Sect.~\ref{sec:results:general}.
As expected, our sample is less complete towards the fainter end of the optical and X-ray fluxes.
For X-ray sources with a very reliable counterpart (\texttt{CTP\_QUALITY}$>1$) alone, with a highly significant X-ray detection (\texttt{DET\_LIKE}$>10$), or within the footprints of the SDSS observed plates (DEC$>2^{\circ}$), we defined a cleaner sample for which the spectroscopic completeness reached 81\% for $r < 21.38$.
The completeness decreased markedly below this limit.
To observe significantly fainter objects with similar success rates, the next generation of large area surveys that are conducted with 4m telescopes such as DESI \citep{DESI2022} and 4MOST \citep{4MOST2019} are required. 

From the qualitative analysis of the stacked spectra we presented in Sect.~\ref{sec:results:xray_pointlike}, the diversity of the spectral types of the X-ray selected AGN in the eFEDS field becomes clear.
The examples showed obscured (narrow-line galaxies) and unobscured AGNs, with different fractions of contribution from the host galaxy and from blue and red quasars. 
Other types of spectra, such as those from passive galaxies without emission lines, will not be evident through the stacking technique when they are mixed with spectra that present emission lines.
These cases can be found after a visual inspection of the eFEDS individual spectra, as in the case of the AGNs within clusters of galaxies (\texttt{CLUSTER\_CLASS$\geq4$}). 

The large statistical sample presented here shows that robust trends connect the optical colour and the average X-ray column density in a given redshift interval.
The sample diversity indicates its potential for the study of AGN demography and the evolution of AGNs across cosmic time through an individual study of each object or through population studies with approaches similar to those presented here.

Faithful to its nature as a pathfinder (and performance verification), the eFEDS survey paves the way to a much larger and ambitious spectroscopic follow-up program in SDSS-V BHM (Kollmeier et al., in prep.).
This aims to obtain about 400k optical spectra by 2027 with a similar quality as those presented here for the X-ray sources that are detected in the eROSITA all-sky survey.

Future work (Aydar et al. in prep) will present the optical spectral fitting of each object in eFEDS.
We will provide measurements of the continuum and line features to estimate the physical properties of AGNs such as the black hole mass, the host-galaxy properties and their contribution to the observed emission, the AGN classification, the presence of outflows, and the study of specific emission lines such as coronal lines (i.e. emission lines with an ionisation potential $\geq100$ eV).
Combined with those obtained from the X-ray observations (e.g. X-ray luminosity, column density, and photon index), these properties will provide a better understanding of the role of obscuration in the AGN classification, the setting of possible calibrations for scaling relations and black hole mass estimates, and a large and uniform sample that provides reliable statistics of the AGN populations according to different physical parameters.

\section{Data availability}
Table eFEDS\_VAC143\_AGN181\_CTP17\_PipelineRedshift.fits described in the Appendix \ref{sec:appendix:catalogues} and Table spectra\_compilation\_eFEDS\_v1.4.3.fits described in the Appendix \ref{sec:data_model_efeds_speccomp} are available in electronic form at the CDS via anonymous ftp to cdsarc.u-strasbg.fr (130.79.128.5) or via \url{http://cdsweb.u-strasbg.fr/cgi-bin/qcat?J/A+A/}.
These catalogues are also available at the \href{https://erosita.mpe.mpg.de/edr/eROSITAObservations/Catalogues/}{eROSITA-DE Early Data Release website}.

\begin{acknowledgements}
This work is based on data from eROSITA, the primary instrument aboard SRG, a joint Russian-German science mission supported by the Russian Space Agency (Roskosmos), in the interests of the Russian Academy of Sciences represented by its Space Research Institute (IKI), and the Deutsches Zentrum f\"ur Luft- und Raumfahrt (DLR). The SRG spacecraft was built by Lavochkin Association (NPOL) and its subcontractors, and is operated by NPOL with support from the Max Planck Institute for Extraterrestrial Physics (MPE).

The development and construction of the eROSITA X-ray instrument was led by MPE, with contributions from the Dr. Karl Remeis Observatory Bamberg \& ECAP (FAU Erlangen-N\"urnberg), the University of Hamburg Observatory, the Leibniz Institute for Astrophysics Potsdam (AIP), and the Institute for Astronomy and Astrophysics of the University of T\"ubingen, with the support of DLR and the Max Planck Society. The Argelander Institute for Astronomy of the University of Bonn and the Ludwig Maximilians Universit\"at Munich also participated in the science preparation for eROSITA. The eROSITA data shown here were processed using the eSASS software system developed by the German eROSITA consortium.

Funding for the Sloan Digital Sky Survey V has been provided by the Alfred P. Sloan Foundation, the Heising-Simons Foundation, the National Science Foundation, and the Participating Institutions. SDSS acknowledges support and resources from the Center for High-Performance Computing at the University of Utah. The SDSS web site is \url{www.sdss.org}.

SDSS is managed by the Astrophysical Research Consortium for the Participating Institutions of the SDSS Collaboration, including the Carnegie Institution for Science, Chilean National Time Allocation Committee (CNTAC) ratified researchers, Caltech, the Gotham Participation Group, Harvard University, Heidelberg University, The Flatiron Institute, The Johns Hopkins University, L'Ecole polytechnique f\'{e}d\'{e}rale de Lausanne (EPFL), Leibniz-Institut f\"{u}r Astrophysik Potsdam (AIP), Max-Planck-Institut f\"{u}r Astronomie (MPIA Heidelberg), Max-Planck-Institut f\"{u}r Extraterrestrische Physik (MPE), Nanjing University, National Astronomical Observatories of China (NAOC), New Mexico State University, The Ohio State University, Pennsylvania State University, Smithsonian Astrophysical Observatory, Space Telescope Science Institute (STScI), the Stellar Astrophysics Participation Group, Universidad Nacional Aut\'{o}noma de M\'{e}xico, University of Arizona, University of Colorado Boulder, University of Illinois at Urbana-Champaign, University of Toronto, University of Utah, University of Virginia, Yale University, and Yunnan University.

CA acknowledges the support of the Excellence Cluster ORIGINS, which is funded by the Deutsche Forschungsgemeinschaft (DFG, German Research Foundation) under Germany's Excellence Strategy – EXC-2094 – 390783311.

FEB acknowledges support from ANID-Chile BASAL CATA FB210003, FONDECYT Regular 1241005, and Millennium Science Initiative, AIM23-0001.

MB acknowledges support from the European Union’s Innovative Training Network (ITN) funded by the Marie Sklodowska-Curie Actions in Horizon 2020 No 860744 (BiD4BEST).  

WB acknowledges the Eberly Endowment at Penn State.

YD acknowledges financial support from a Fondecyt postdoctoral fellowship (3230310).

LHG acknowledges funding from ANID programs: FONDECYT Iniciaci\'on 11241477,  Millennium Science Initiative Program NCN$2023\_002$, and Millennium Science Initiative AIM23-0001.

TM is supported by UNAM-DGAPA by PAPIIT IN114423. TM also thanks JSPS for its support under the  Invitational Fellowships for Research in Japan program (L-24523) during his sabbatical stay at Kyoto University.

CAN and HJIM acknowledge the support from projects CONAHCyT CBF2023-2024-1418, PAPIIT IA104325 and IN119123.

CR acknowledges support from Fondecyt Regular grant 1230345, ANID BASAL project FB210003 and the China-Chile joint research fund.

RA was supported by NASA through the NASA Hubble Fellowship grant \#HST-HF2-51499.001-A awarded by the Space Telescope Science Institute, which is operated by the Association of Universities for Research in Astronomy, Incorporated, under NASA contract NAS5-26555.

MK is supported by DLR grant FKZ 50 OR 2307.

\end{acknowledgements}

\bibliographystyle{aa}
\bibliography{eFEDScatbib}

\begin{appendix}
\onecolumn

\section{Description of released datasets}
\label{sec:data_format}

In this section, we describe the contents and format of the data files released alongside this work.

\subsection{Catalogues}
\label{sec:appendix:catalogues}
The main catalogues used for this paper are shown in Table \ref{tab:catalogs}, including those released here for the first time.
The file eFEDS\_VAC143\_AGN181\_CTP17\_PipelineRedshift.fits is a compilation of the works by \citet{Brunner2022}, \citet{Salvato2022}, \citet{Liu2022}, and \citet{Almeida2023_sdss_dr18} with the data that was considered for the plots in this manuscript.

\begin{table*}[h]
\caption{File name, number of entries, and reference for the catalogues that were used in this paper.}
\label{tab:catalogs}
\centering
\begin{tabular}{lll}
\hline\hline 
Catalogue  & Number of entries & Reference \\
\hline
\href{https://www.sdss.org/dr18/data\_access/value-added-catalogs/?vac\_id=10001}{eFEDS\_SDSSV\_spec\_results-v1.4.3.fits}                  & 13085          & \citet{Almeida2023_sdss_dr18} and Table \ref{tab:efeds_sdssv_spec_results} \\ 
\href{}{spectra\_compilation\_eFEDS\_v1.4.3.fits}                                                       & 334258                & This work and Tables \ref{tab:efeds_specz_compilation_datamodel_a}-\ref{tab:efeds_specz_compilation_datamodel_b} \\ 
\href{https://www.sdss.org/dr18/data_access/value-added-catalogs/?vac_id=10001}{eFEDS\_Main\_speccomp-v1.4.3.fits}  & 27369           & \citet{Almeida2023_sdss_dr18} and Tables \ref{tab:efeds_main_specomp_datamodel_a}-\ref{tab:efeds_main_specomp_datamodel_b}  \\ 
\href{https://erosita.mpe.mpg.de/edr/eROSITAObservations/Catalogues/}{eFEDS\_c001\_main\_V7.4.fits} & 27910 & \citet{Brunner2022} \\
\href{https://erosita.mpe.mpg.de/edr/eROSITAObservations/Catalogues/}{eFEDS\_C001\_Main\_PointSources\_CTP\_redshift\_V17.fits} & 27369             & \citet{Salvato2022}             \\ 
\href{https://erosita.mpe.mpg.de/edr/eROSITAObservations/Catalogues/liuT/eFEDS_AGN_spec_V10.html}{eFEDS\_AGN\_spec\_V18.1.fits}                             & 27910             & \citet{Liu2022}, see footnote 6 \\  
\href{}{eFEDS\_VAC143\_AGN181\_CTP17\_PipelineRedshift.fits} & 27369 & \makecell[l]{This work, based on the previously \\released catalogs} \\ 
\hline
\end{tabular}
\end{table*}

\subsection{SDSS-V spectroscopy obtained in the eFEDS field, with visual inspection information}
\label{sec:data_model_sdssv_spectroscopy}
This data table was originally released as part of SDSS DR18 \citep{Almeida2023_sdss_dr18}, but was not fully described. 
Therefore, in Table \ref{tab:efeds_sdssv_spec_results}, we provide a complete column-by-column description.

\begin{table*}[h]
\caption{Data model for the SDSS-V/eFEDS dataset - a catalogue of spectra obtained during the SDSS-V/eFEDS campaign, including visual inspection information.}
\label{tab:efeds_sdssv_spec_results}
\centering
\begin{tabular}{llp{25em}l}
    \hline\hline 
      Column name & Datatype & Description & Units \\
     \hline
        field & smallint & SDSS field code identifier & ~ \\ 
        mjd & int & SDSS MJD associated with this spectrum & ~ \\ 
        catalogid & long & SDSS-V CATALOGID (v0) associated with this target & ~ \\ 
        plug\_ra & float & Sky coordinate of spectroscopic fibre & deg \\ 
        plug\_dec & float & Sky coordinate of spectroscopic fibre & deg \\ 
        nvi & int & Number of visual inspections collected for this spectrum & ~ \\ 
        sn\_median\_all & float & Median S/N per pix in spectrum (idlspec2d eFEDS v6\_0\_2 reductions) & ~ \\ 
        z\_pipe & float & Pipeline redshift in idlspec1d eFEDS v6\_0\_2 reductions & ~ \\ 
        z\_err\_pipe & float & Pipeline redshift uncertainty in idlspec1d eFEDS v6\_0\_2 reductions & ~ \\ 
        zwarning\_pipe & int & Pipeline redshift warning flags in idlspec1d eFEDS v6\_0\_2 reductions & ~ \\ 
        class\_pipe & string & Pipeline classification in idlspec1d eFEDS v6\_0\_2 reductions & ~ \\ 
        subclass\_pipe & string & Pipeline sub-classification in idlspec1d eFEDS v6\_0\_2 reductions & ~ \\ 
        z\_final & float & Final redshift derived from pipeline and visual inspections & ~ \\ 
        z\_conf\_final & int & Final redshift confidence from the pipeline and visual inspections & ~ \\ 
        class\_final & string & Final classification derived from pipeline and visual inspections & ~ \\ 
        blazar\_candidate & Boolean & Was the object flagged as a blazar candidate in visual inspections? & ~ \\ 
\end{tabular}
\end{table*}

\subsection{The eFEDS spectroscopic compilation (all spec-z)}
\label{sec:data_model_efeds_speccomp}
\textit{Notes on content and usage.} 
This catalogue contains a row for each pairing of an original spec-z and an object from 
the DESI Legacy Imaging Survey DR9 optical-to-IR catalogue \citep[][LS9]{Dey2019}. 
The algorithm deliberately allows a single spec-z to be associated with multiple LS9 objects.
In some cases, this is because the spectroscopic aperture covers more than one astrophysical object, but in other
cases, it appears that a single astrophysical object has been split into two or more entries in the LS9 catalogue. 
Note, selecting entries with \texttt{ORIG\_LS\_RANK\_CTP} = 1 gives only the nearest LS9 object (to the \texttt{SPECZ\_RA}, \texttt{SPECZ\_DEC} coordinates). 
Selecting rows with \texttt{SPECZ\_RANK} = 1 gives only a single best REDSHIFT, CLASS, NORMQ per astrophysical object, which is generally what is desired for science analyses. 
Selecting rows with \texttt{SPECZ\_NORMQ} = 3 retains only the spectroscopic redshifts that we consider to have the highest reliability and quality.
In Tables \ref{tab:efeds_specz_compilation_datamodel_a}-\ref{tab:efeds_specz_compilation_datamodel_b} we describe the column by column content.

\begin{longtable}{llp{25em}l}
\caption{Data model for the eFEDS spectroscopic compilation (all spec-z), part 1 of 2.} 
\label{tab:efeds_specz_compilation_datamodel_a} \\
\hline\hline 
Column name & Datatype & Description & Units \\
\hline
\endfirsthead
\caption[]{continued.} \\
\hline\hline
Column name & Datatype & Description & Units \\
\hline
\endhead
\hline
\endfoot
        ls\_idx & long & Index for internal use only & ~ \\
        ls\_id & long & Unique ID of Legacy Survey DR9 photometric object associated with the spec-z. Computed from Legacy Survey DR9 catalogue as ls\_id = objid + (brickid $<<$ 16) + (release $<<$ 40) & ~ \\ 
        ls\_ra & float & Coordinate from Legacy Survey DR9 at epoch ls\_epoch & deg \\ 
        ls\_dec & float & Coordinate from Legacy Survey DR9 at epoch ls\_epoch & deg \\ 
        ls\_pmra & float & Proper motion from Legacy Survey DR9 & mas~yr$^{-1}$ \\ 
        ls\_pmdec & float & Proper motion from Legacy Survey DR9 & mas~yr$^{-1}$ \\ 
        ls\_epoch & float & Coordinate epoch from Legacy Survey DR9 & year \\ 
        ls\_mag\_g & float & DECam g-band model magnitude derived from Legacy Survey DR9, AB & mag \\ 
        ls\_mag\_r & float & DECam r-band model magnitude derived from Legacy Survey DR9, AB & mag \\ 
        ls\_mag\_z & float & DECam z-band model magnitude derived from Legacy Survey DR9, AB & mag \\ 
        specz\_n & int & Total number of spec-z associated with this Legacy Survey DR9 object & ~ \\ 
        specz\_raj2000 & float & Coordinate of spec-z, propagated if necessary to epoch J2000 & deg \\ 
        specz\_dej2000 & float & Coordinate of spec-z, propagated if necessary to epoch J2000 & deg \\ 
        specz\_nsel & int & Number of spec-z selected to inform result for this object & ~ \\ 
        specz\_rank & int & Rank of this particular spec-z (1=best, >1=less favoured, 0=not selected) & ~ \\  
        specz\_redshift & float & Final redshift determined for this object & ~ \\ 
        specz\_normq & int & Final normalised redshift quality (\texttt{NORMQ}) associated with this object (see section \ref{sec:methods}) & ~ \\ 
        specz\_normc & string & Final normalised classification determined for this object & ~ \\ 
        specz\_hasvi & Boolean & True if best spec-z for this object has a visual inspection & ~ \\ 
        specz\_catcode & string & Catalogue code of best spec-z for this object (see table \ref{tab:specz_comp:inputs}) & ~ \\ 
        specz\_bitmask & long & Bitmask encoding catalogues containing spec-z for this object. Bit encoding: 0=sdssv\_vi; 1=boss\_vi; 2=sdss\_vi; 3=efeds\_vi; 4=gama; 5=wigglez; 6=2slaq; 7=6dfgs; 8=2mrs; 9=boss\_novi; 10=sdss\_novi; 11=hectospec; 12=fast; 13=gaia\_rvs; 14=lamost; 15=simbad; 16=ned & ~ \\ 
        specz\_sel\_bitmask & long & Bitmask encoding catalogues containing informative spec-z for an object. Bit encoding is the same as specz\_bitmask  & ~ \\ 
        specz\_flags & int & Bitmask encoding quality flags for this object. Bit encoding: 0=No Legacy Survey DR9 counterpart within matching radius; 1=Significant scatter ($\mathrm{stddev}(z_i)>0.01$) in the redshifts available for this object; 2=unused; 3=Blazar candidate; 4=Disagreement between the normalised classifications available for this object & ~ \\ 
        specz\_sel\_normq\_max & int & Highest \texttt{NORMQ} of informative spec-z for this object & ~ \\ 
        specz\_sel\_normq\_mean & float & Mean \texttt{NORMQ} of informative spec-z for this object & ~ \\ 
        specz\_sel\_z\_mean & float & Mean redshift of informative spec-z for this object & ~ \\ 
        specz\_sel\_z\_median & float & Median redshift of informative spec-z for this object & ~ \\ 
        specz\_sel\_z\_stddev & float & Standard deviation of redshifts for informative spec-z for object & ~ \\
        specz\_all\_nvi & int & Number of spec-z with VIs for this object & ~ \\
        specz\_all\_normq\_max & float & Highest NORMQ of all spec-z for this object & ~ \\
        specz\_all\_normq\_mean & float & Mean NORMQ of all spec-z for this object & ~ \\
        specz\_all\_z\_mean & float & Mean redshift of all spec-z for this object & ~ \\
        specz\_all\_z\_median & float & Median redshift of all spec-z for this object & ~ \\
        specz\_all\_z\_stddev & float & Standard deviation of redshifts for all spec-z for this object & ~ \\
\end{longtable}

\begin{longtable}{llp{25em}l}
\caption{Data model for the eFEDS spectroscopic compilation (all spec-z), part 2 of 2.}
\label{tab:efeds_specz_compilation_datamodel_b} \\
\hline\hline 
Column name & Datatype & Description & Units \\
\hline
\endfirsthead
\caption[]{continued.} \\
\hline\hline
Column name & Datatype & Description & Units \\
\hline
\endhead
\hline
\endfoot
        orig\_catcode & string & Catalogue code of individual spec-z & ~ \\ 
        orig\_ra & float & Coordinate associated with individual spec-z measurement & deg \\ 
        orig\_dec & float & Coordinate associated with individual spec-z measurement & deg \\ 
        orig\_pos\_epoch & float & Coordinate epoch associated with individual spec-z measurement & ~ \\ 
        orig\_ls\_sep & float & Distance from spec-z to Legacy Survey DR9 photometric counterpart (corrected for proper motion) & arcsec \\ 
        orig\_ls\_gt1ctp & Boolean & Can spec-z be associated with $>$1 possible Legacy Survey DR9 counterpart? & ~ \\ 
        orig\_ls\_ctp\_rank & int & Rank of counterpart out of all possibilities for this spec-z (1=closest) & ~ \\ 
        orig\_id\_col & string & Name of column in original catalogue supplying ORIG\_ID value & ~ \\ 
        orig\_qual\_col & string & Name of column in original catalogue supplying ORIG\_QUAL value & ~ \\ 
        orig\_redshift\_col & string & Name of column in original catalogue supplying ORIG\_REDSHIFT value & ~ \\ 
        orig\_class\_col & string & Name of column in original catalogue supplying ORIG\_CLASS value & ~ \\ 
        orig\_id & string & Orig. value of ID of individual spec-z measurement (as a string) & ~ \\ 
        orig\_redshift & float & Orig. redshift value of individual spec-z measurement & ~ \\ 
        orig\_qual & string & Orig. redshift quality value of individual spec-z measurement & ~ \\ 
        orig\_normq & int & Normalised redshift quality of individual spec-z measurement & ~ \\ 
        orig\_class & string & Orig. classification label of individual spec-z measurement & ~ \\ 
        orig\_hasvi & Boolean & True if individual spec-z has a visual inspection from our team & ~ \\ 
        orig\_normc & string & Normalised classification code of individual spec-z measurement & ~ \\ 
        orig\_bitmask & long & Bitmask encoding catalogue providing spec-z for this spec-z measurement. See specz\_bitmask column. & ~ \\ 
\end{longtable}

\subsection{The eFEDS Main sample catalogue with spectroscopic information}
\label{sec:data_model_efeds_main_with_speccomp}
This catalogue table is presented in section \ref{sec:efeds_main_cat_with_specz}. 
The data table was originally released as part of SDSS DR18 \citep{Almeida2023_sdss_dr18}, but was not fully described. Therefore, in Tables \ref{tab:efeds_main_specomp_datamodel_a}-\ref{tab:efeds_main_specomp_datamodel_b}, we provide a complete column-by-column description.

\newpage
\begin{longtable}{llp{25em}l}
\caption{Data model for the eFEDS Main X-ray counterpart catalogue combined with the spectroscopic compilation, part 1 of 2.} 
\label{tab:efeds_main_specomp_datamodel_a} \\
\hline\hline 
Column name & Datatype & Description & Units \\
\hline
\endfirsthead
\caption[]{continued.} \\
\hline\hline
Column name & Datatype & Description & Units \\
\hline
\endhead
\hline
\endfoot
        ero\_name & string & From \citet{Brunner2022}, eROSITA official source Name & ~ \\ 
        ero\_id\_src & int & From \citet{Brunner2022}, ID of eROSITA source in the Main Sample & ~ \\ 
        ero\_ra\_corr & float & From \citet{Brunner2022}, J2000 Right Ascension of eROSITA source (corrected) & deg \\ 
        ero\_dec\_corr & float & From \citet{Brunner2022}, J2000 Declination of eROSITA source (corrected) & deg \\ 
        ero\_radec\_err\_corr & float & From \citet{Brunner2022}, eROSITA positional uncertainty (corrected) & arcsec \\ 
        ero\_ml\_flux & float & From \citet{Brunner2022}, 0.2-2.3 keV source flux & erg~cm$^{-2}$~s$^{-1}$ \\ 
        ero\_ml\_flux\_err & float & From \citet{Brunner2022}, 0.2-2.3 keV source flux error (1 sigma) & erg~cm$^{-2}$~s$^{-1}$ \\ 
        ero\_det\_like & float & From \citet{Brunner2022}, 0.2-2.3 keV detection likelihood via PSF-fitting & ~ \\ 
        ctp\_ls8\_unique\_objid & string & From \citet{Salvato2022},  unique id for Legacy Survey DR8 counterpart to the eROSITA source & ~ \\ 
        ctp\_ls8\_ra & float & From \citet{Salvato2022}, Right Ascension of the Legacy Survey DR8 counterpart & deg \\ 
        ctp\_ls8\_dec & float & From \citet{Salvato2022}, Declination of the best Legacy Survey DR8 counterpart & deg \\ 
        dist\_ctp\_ls8\_ero & float & From \citet{Salvato2022}, Separation between counterpart and eROSITA position & arcsec \\ 
        ctp\_quality & smallint & From \citet{Salvato2022}, counterpart quality: 4=best, 3=good, 2=secondary, 1 or 0=unreliable & ~ \\ 
        ls\_id & long & Unique ID of Legacy Survey DR9 photometric object associated with the spec-z. Computed from Legacy Survey DR9 catalogue as ls\_id = objid + (brickid $<<$ 16) + (release $<<$ 40) & ~ \\ 
        ls\_ra & float & Coordinate from Legacy Survey DR9 at epoch ls\_epoch & deg \\ 
        ls\_dec & float & Coordinate from Legacy Survey DR9 at epoch ls\_epoch & deg \\ 
        ls\_pmra & float & Proper motion from Legacy Survey DR9 & mas~yr$^{-1}$ \\ 
        ls\_pmdec & float & Proper motion from Legacy Survey DR9 & mas~yr$^{-1}$ \\ 
        ls\_epoch & float & Coordinate epoch from Legacy Survey DR9 & year \\ 
        ls\_mag\_g & float & DECam g-band model magnitude derived from Legacy Survey DR9, AB & mag \\ 
        ls\_mag\_r & float & DECam r-band model magnitude derived from Legacy Survey DR9, AB & mag \\ 
        ls\_mag\_z & float & DECam z-band model magnitude derived from Legacy Survey DR9, AB & mag \\ 
\end{longtable}

\newpage
\begin{longtable}{llp{25em}l}
\caption{Data model for the eFEDS Main X-ray counterpart catalogue combined with the spectroscopic compilation, part 2 of 2.}
\label{tab:efeds_main_specomp_datamodel_b} \\
\hline\hline 
Column name & Datatype & Description & Units \\
\hline
\endfirsthead
\caption[]{continued.} \\
\hline\hline
Column name & Datatype & Description & Units \\
\hline
\endhead
\hline
\endfoot        
        specz\_n & int & Total number of spec-z associated with this Legacy Survey DR9 object & ~ \\ 
        specz\_raj2000 & float & Coordinate of spec-z, propagated if necessary to epoch J2000 & deg \\ 
        specz\_dej2000 & float & Coordinate of spec-z, propagated if necessary to epoch J2000 & deg \\ 
        specz\_nsel & int & Number of spec-z selected to inform result for this object & ~ \\ 
        specz\_redshift & float & Final redshift determined for this object & ~ \\ 
        specz\_normq & int & Final normalised redshift quality (\texttt{NORMQ}) associated with this object (see section \ref{sec:methods}) & ~ \\ 
        specz\_normc & string & Final normalised classification determined for this object & ~ \\ 
        specz\_hasvi & Boolean & True if best spec-z for this object has a visual inspection & ~ \\ 
        specz\_catcode & string & Catalogue code of best spec-z for this object (see table \ref{tab:specz_comp:inputs}) & ~ \\ 
        specz\_bitmask & long & Bitmask encoding catalogues containing spec-z for this object. Bit encoding: 0=sdssv\_vi; 1=boss\_vi; 2=sdss\_vi; 3=efeds\_vi; 4=gama; 5=wigglez; 6=2slaq; 7=6dfgs; 8=2mrs; 9=boss\_novi; 10=sdss\_novi; 11=hectospec; 12=fast; 13=gaia\_rvs; 14=lamost; 15=simbad; 16=ned & ~ \\ 
        specz\_sel\_bitmask & long & Bitmask encoding catalogues containing informative spec-z for an object. Bit encoding is the same as specz\_bitmask  & ~ \\ 
        specz\_flags & int & Bitmask encoding quality flags for this object. Bit encoding: 0=No Legacy Survey DR9 counterpart within matching radius; 1=Significant scatter ($\mathrm{stddev}(z_i)>0.01$) in the redshifts available for this object; 2=unused; 3=Blazar candidate; 4=Disagreement between the normalised classifications available for this object & ~ \\ 
        specz\_sel\_normq\_max & int & Highest \texttt{NORMQ} of informative spec-z for this object & ~ \\ 
        specz\_sel\_normq\_mean & float & Mean \texttt{NORMQ} of informative spec-z for this object & ~ \\ 
        specz\_sel\_z\_mean & float & Mean redshift of informative spec-z for this object & ~ \\ 
        specz\_sel\_z\_median & float & Median redshift of informative spec-z for this object & ~ \\ 
        specz\_sel\_z\_stddev & float & Standard deviation of redshifts for informative spec-z for object & ~ \\ 
        specz\_orig\_ra & float & Coordinate associated with individual spec-z measurement & deg \\ 
        specz\_orig\_dec & float & Coordinate associated with individual spec-z measurement & deg \\ 
        specz\_orig\_pos\_epoch & float & Coordinate epoch associated with individual spec-z measurement & ~ \\ 
        specz\_orig\_ls\_sep & float & Distance from spec-z to Legacy Survey DR9 photometric counterpart (corrected for proper motion) & arcsec \\ 
        specz\_orig\_ls\_gt1ctp & Boolean & Can spec-z be associated with $>$1 possible Legacy Survey DR9 counterpart? & ~ \\ 
        specz\_orig\_ls\_ctp\_rank & int & Rank of counterpart out of all possibilities for this spec-z (1=closest) & ~ \\ 
        specz\_orig\_id & string & Orig. value of ID of individual spec-z measurement (as a string) & ~ \\ 
        specz\_orig\_redshift & float & Orig. redshift value of individual spec-z measurement & ~ \\ 
        specz\_orig\_qual & string & Orig. redshift quality value of individual spec-z measurement & ~ \\ 
        specz\_orig\_normq & int & Normalised redshift quality of individual spec-z measurement & ~ \\ 
        specz\_orig\_class & string & Orig. classification label of individual spec-z measurement & ~ \\ 
        specz\_orig\_hasvi & Boolean & True if individual spec-z has a visual inspection from our team & ~ \\ 
        specz\_orig\_normc & string & Normalised classification code of individual spec-z measurement & ~ \\ 
        specz\_ra\_used & float & Adopted coordinate of specz when matching to \citet{Salvato2022} counterpart & deg \\ 
        specz\_dec\_used & float & Adopted coordinate of specz when matching to \citet{Salvato2022} counterpart & deg \\ 
        separation\_specz\_ctp & float & Distance from LS\_RA,LS\_DEC to SPECZ\_RA\_USED,SPECZ\_DEC\_USED & arcsec \\ 
        has\_specz & Boolean & Does this \citet{Salvato2022} counterpart have a spec-z? & ~ \\ 
        has\_informative\_specz & Boolean & Does this \citet{Salvato2022} counterpart have an informative spec-z? & ~ \\ 
\end{longtable}

\end{appendix}

\end{document}